\documentclass[epj]{svjour}
\usepackage[T1]{fontenc}
\usepackage[latin1]{inputenc}
\usepackage{graphicx}
\usepackage{amssymb}

\makeatletter

\usepackage{color}

\usepackage{multibox}

\def\tensor#1{{\vec{#1}}}
\def\ellperso{\vec{\ell}}
\def\ellperso{\mbox{\boldmath $\ell$}}
\def\eps{\mbox{\boldmath $\varepsilon$}}

\def\deps{{\dot{\mbox{\boldmath $\varepsilon$}}}}
\def\scalarstrain{ U_d}
\def\scalarstrainsquared{ (U_d)^2}
\def\tensorstrain{\tensor{U}}

\makeatletter
\usepackage{epstopdf}
\graphicspath{{analyseImagesMariusGeorges/Marius/},{analyseImagesMariusGeorges/Georges/},{T1Christophe/},{aFigures/}}
\usepackage{umlaut}

\makeatother

\makeatother
\begin{document}
\title{Discrete rearranging disordered patterns, part II: 2D plasticity, elasticity and flow of a foam}
\author{P. Marmottant
\thanks{Author for correspondence at philippe.marmottant@ujf-grenoble.fr}
\and C. Raufaste
\and F. Graner}
\institute{Laboratoire de Spectrom\'{e}trie Physique, UMR5588, CNRS-Universit\'{e} Grenoble I, B.P. 87, F-38402 St Martin
d'H\`{e}res Cedex, France}
\date{\today}
\abstract{
The plastic flow of a foam results from 
bubble rearrangements.
We study their occurrence in experiments where a foam is forced
to flow in 2D: around an obstacle; through a narrow hole; or sheared between rotating disks. 
We describe their orientation and
frequency using a topological matrix defined in the companion paper  \cite{gra07},
which links them with continuous
plasticity at large scale. 
We then suggest a phenomenological equation to predict the plastic
strain rate: its orientation is determined from the foam's local elastic strain;
and its rate is determined from the foam's local  elongation
rate. 
We obtain a good agreement with statistical measurements. This enables us to describe
the foam as a continuous medium with fluid, elastic and plastic properties. We derive its constitutive equation, then test several of its terms and  predictions. 
} 
\maketitle

\section{Introduction}

\begin{figure}[htb]
(a)
\includegraphics[width=6.5cm,keepaspectratio]{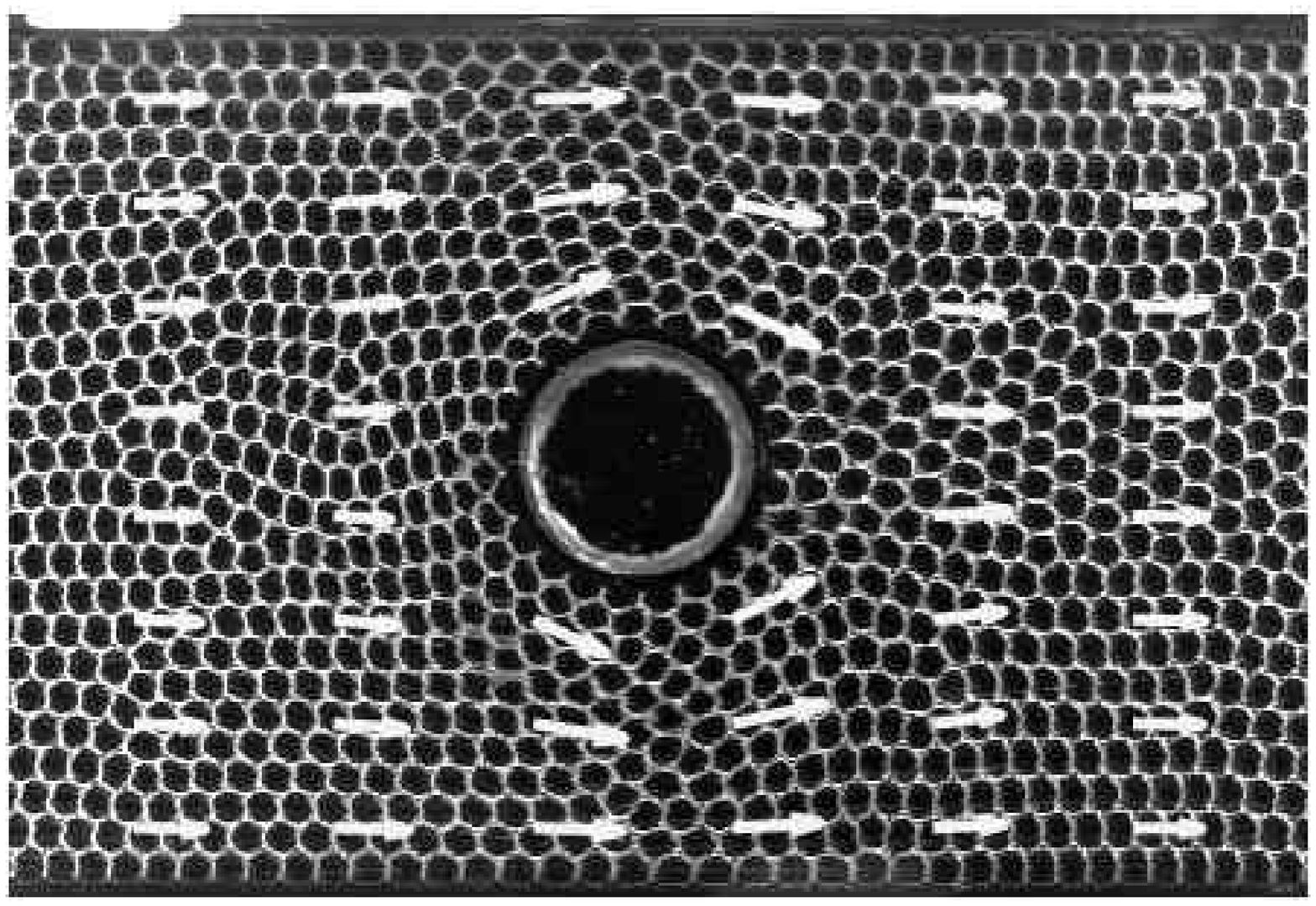}\\ 
(b)
\includegraphics[width=5.76cm,keepaspectratio]{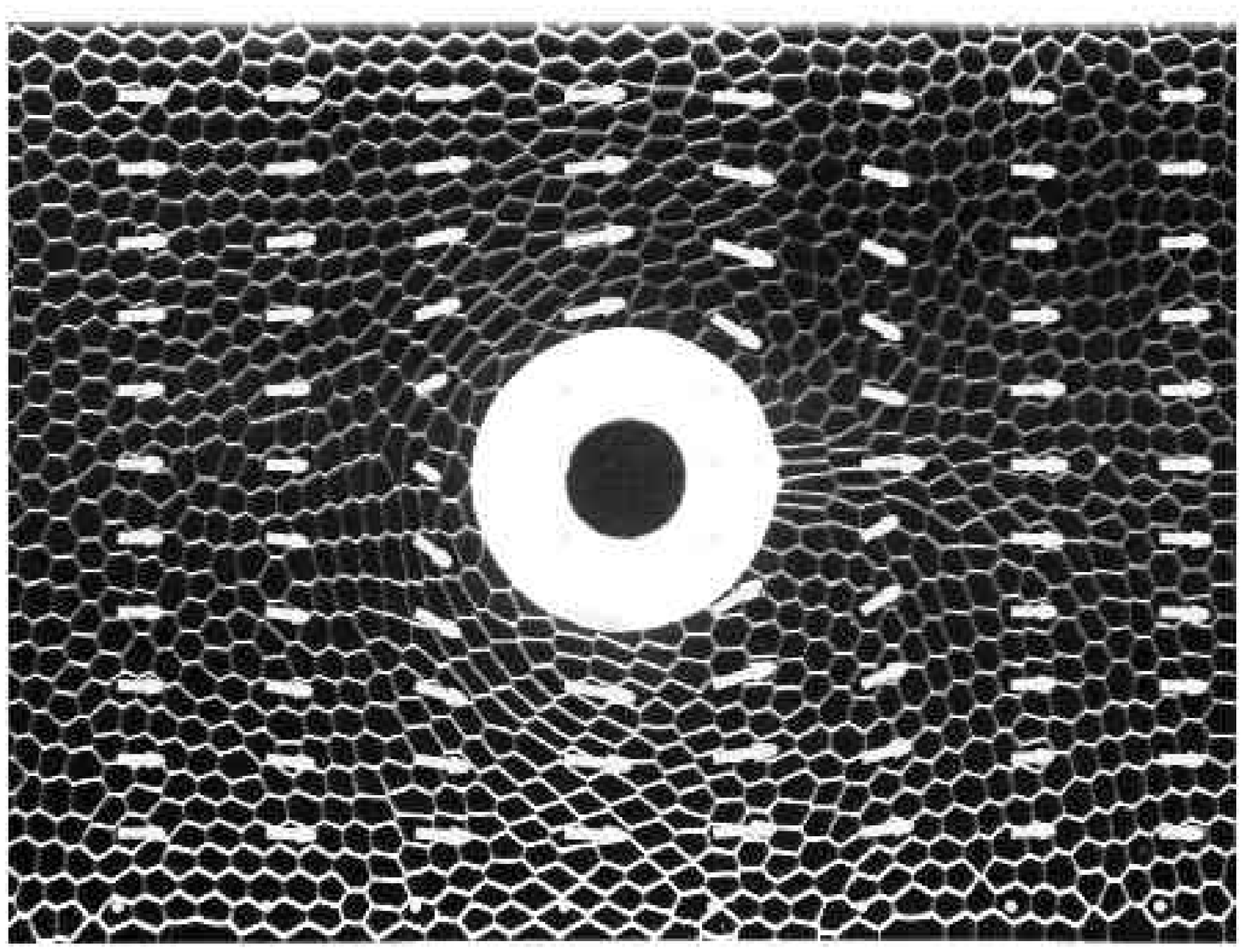}\\
(c)
 \includegraphics[height=3.7cm,angle=90]{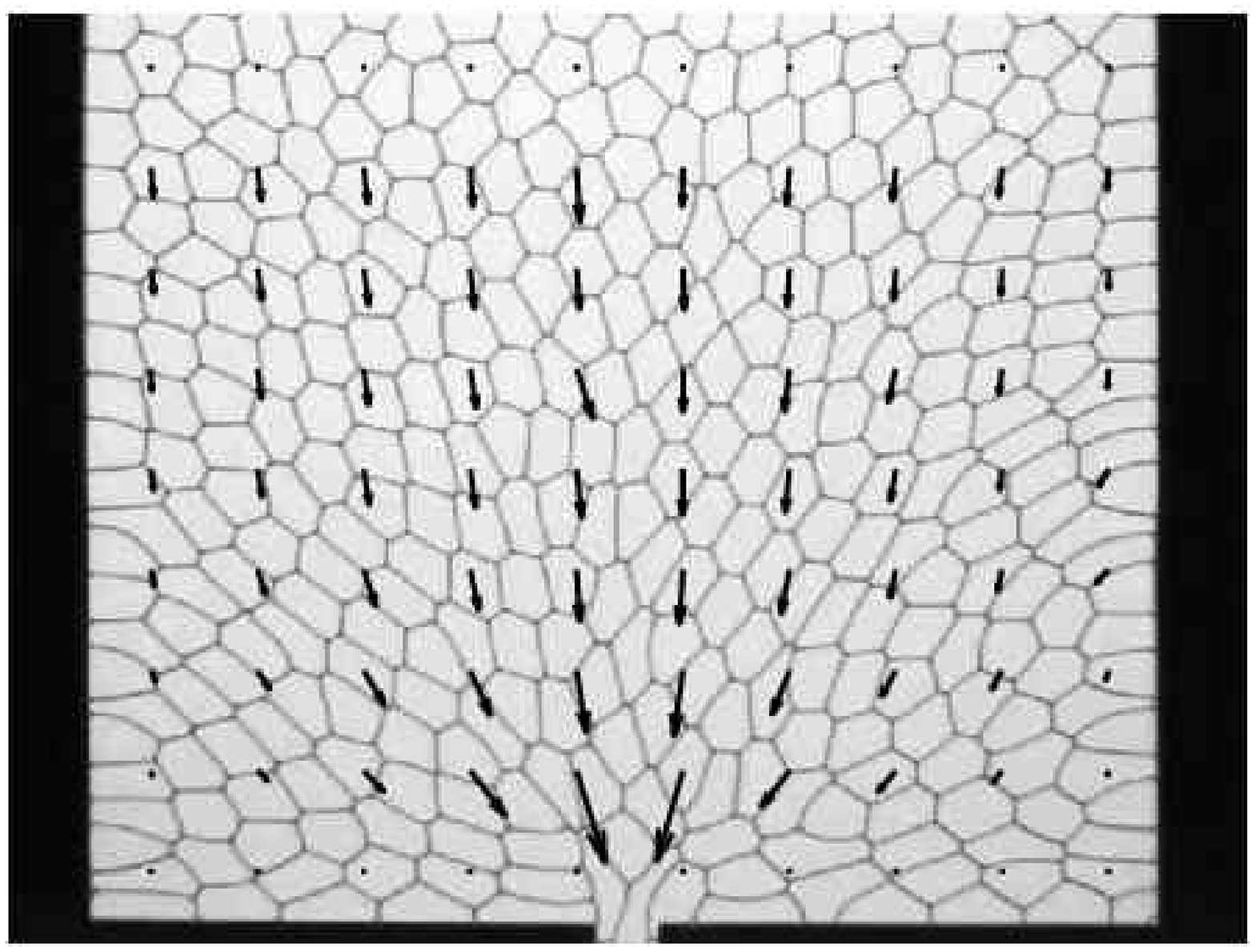} 
(d) 
\includegraphics[width=4cm]{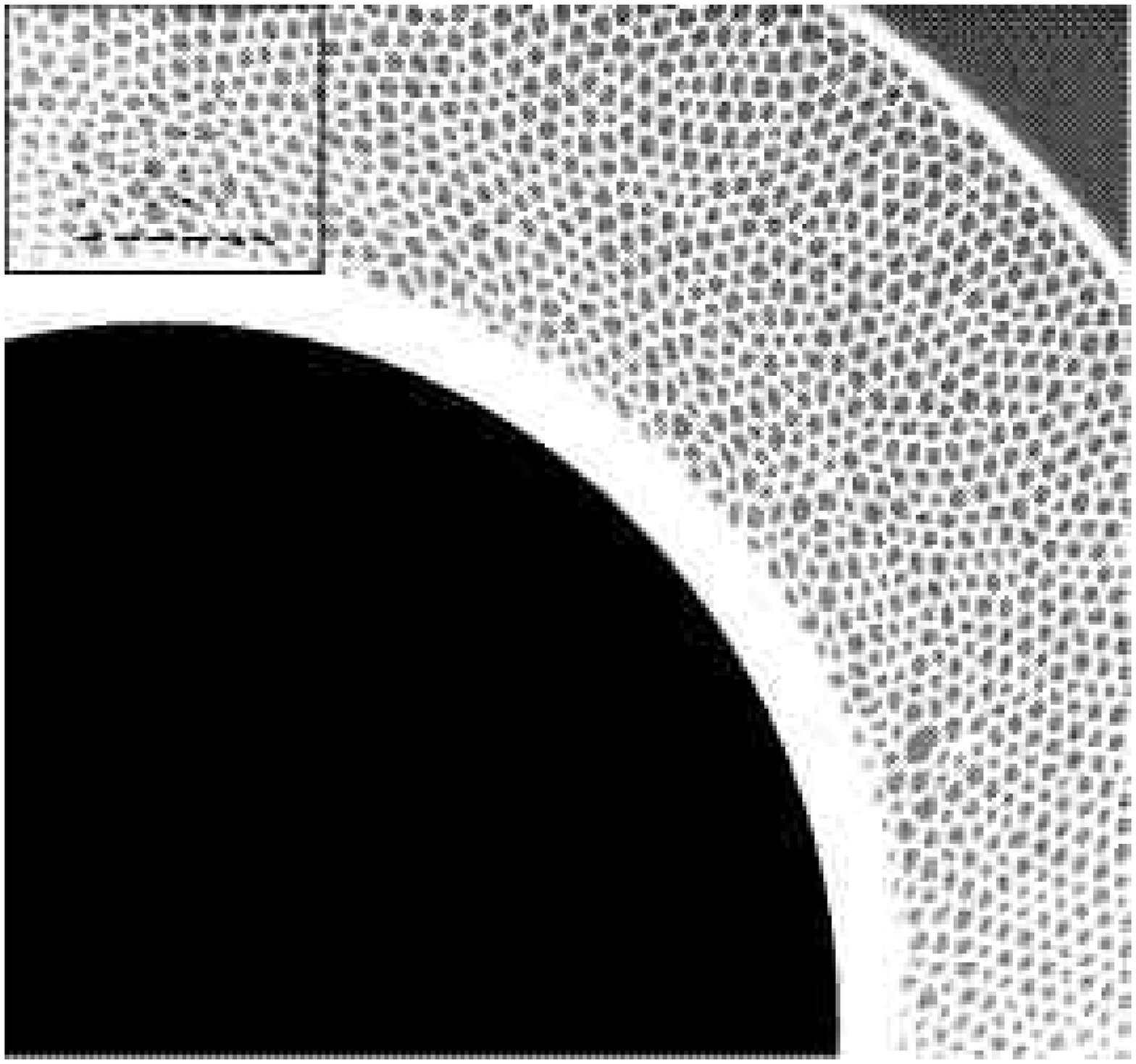}
\caption{
Experimental 2D flows of a foam : top view, here flow from left to right. (a) Wet foam (between glass and water) flowing around an obstacle, picture B. Dollet   \cite{dol05PRE}; field of view 15 cm $\times$ 10 cm. 
(b) Dry foam (between parallel glass plates) flowing around an obstacle,  picture C. Raufaste \cite{gra07}; field of view 13.3 cm $\times$ 10 cm. 
(c) Moderately dry foam (between parallel plexiglass plates) flowing through a narrow hole, picture M. Asipauskas  \cite{asi03}; field of view 7.5 cm $\times$ 10 cm.
(d) Wet foam (between parallel glass plates) sheared between two concentric wheels with tooth-shaped boundaries to prevent slipping; the rotating inner wheel is visible at the bottom, the fixed outer wheel is 
visible at the top \cite{deb01}; arrows indicate the  measured velocity field; ; field of view 
12.5 cm $\times$ 11.5 cm
 Liquid fractions are approximately estimated as: (a) $4\; 10^{-2}$ \cite{Raufaste2007},
(b)  $2\; 10^{-4}$  \cite{rau07}, (c) $10^{-2}$ \cite{asi03}, (d) $5\; 10^{-2}$ \cite{deb01}.
}
\label{setups}
\end{figure}

A liquid foam, made of gas bubbles surrounded by liquid walls (Fig. \ref{setups}), 
is elastic for small strain, plastic for large
strain, and flows at large strain rates \cite{wea99,sai99,hoh05}. 
This complex mechanical behaviour  is exploited in numerous applications,
such as ore separation, drilling and extraction of oil, food or cosmetic
industry \cite{wea99}. It is not yet fully understood \cite{hoh05}. 
Existing models of complex fluids include Oldroyd's visco-elasticity
or Bingham's visco-plasticity \cite{pha02,mac94}. Complete
models tend to unify elastic, plastic and fluid behaviour
 \cite{Schwedoff1900,White1981,jan06,pinceau,Saramito2007,Benito2007preprint,deSouzaMendespreprint}.

Most models describing the foam as a continuous material have to make an assumption to describe phenomenologically the elastic to plastic transition, and the plastic strain rate. Here, we want to link it directly with the observation of foam's individual components, the bubbles. In fact, bubbles play for foams the same role as  microscopical components play for other complex fluids, but have the advantage of being easily observable. The individual plastic events \cite{pic04}  in foams are
the topological rearrangements (neighbour swapping, also called {}"T1
processes\char`\"{} \cite{wea99}). 
 Predicting the occurrence and properties of T1s, and characterizing their impact   at large scale, is thus a crucial step to describe the foam as a continuous material.

Section \ref{ingredients}  lists the ingredients we use. We start from a scalar model of the foam plasticity and dynamics, which predicts correctly its rheological behaviour  \cite{pinceau}. We then include robust statistical tools defined in the companion paper \cite{gra07}, which provides: (i) a link between local (bubble) behaviour and global (foam) measurements of elastic, plastic and fluid behaviours; (ii) matrix (rather than scalar) measurements of these quantities, that is, including direction and magnitude of anisotropy; (iii) space dependence of these measurements  (heterogeneous deformation instead of homogeneous).

Section \ref{plasticmodel} then presents a kinematic equation to describe the elastic to plastic transition,  which is the main point of this paper. It is  an analytical prediction of the plastic strain rate, based on the elastic strain and the total strain rate.

Section \ref{tests} tests its predictions on experiments in different foam flows.  
The relationship between experimental data agrees with our analytical
prediction. Moreover, we correctly predict the frequency and spatial distribution
of T1s, which differs much from the spatial distribution of both
elastic strain and total strain rate.

Section \ref{concl} concludes by  discussing these findings  and   possible applications to other systems. It  proposes a dynamic constitutive equation for foam rheology, relating stresses   to applied strains, with the introduction of a viscosity. 
This closes the complete set of equations describing a foam,  leading to testable predictions.

 

\section{Ingredients of model and tests}

\label{ingredients}
    
\subsection{Continuous description of plasticity with scalar strains}

This section summarizes
 the  model presented in \cite{pinceau} for {\em homogeneous} strains,  described by a simple {\em scalar}. 
This situation arises when considering the strain in a rheometer, where strain is induced by the displacement of a plate with respect to the other, and characterized by a scalar.

The total applied strain rate (or symmetrised velocity gradient) contributes in part to load the elastic  strain (bubbles deform) and to the plastic strain rate (bubbles move relatively to each other). 
The total applied strain rate is the sum of an internal elastic strain rate and the irreversible plastic strain rate: 
 \begin{equation}
\dot{\varepsilon}_{tot} =  \frac{d\varepsilon_{el} }{dt} + \dot{\varepsilon}_{pl},  \label{V_shared_scalar}\\
\end{equation}
where we used a different notation for the time derivative of  $\varepsilon_{el}$, to emphasise that it is an internal state variable of the material.

A   kinematic equation describes how the total strain
rate is shared between change of elastic strain and plastic strain. For slowly sheared foams  
 plasticity occurs if  $ \varepsilon_{el} $ and $ \dot{\varepsilon}_{tot}$ have the same sign (loading of the material), and is zero otherwise  (unloading)
 \cite{pinceau}:
\begin{eqnarray}
\mathrm{if} \quad \varepsilon_{el}  \dot{\varepsilon}_{tot} > 0, \quad
\dot{\varepsilon_{pl}} &=&  h(  \varepsilon_{el} ) \dot{\varepsilon}_{tot}
,\nonumber \\
 \mathrm{if} \quad \varepsilon_{el}  \dot{\varepsilon}_{tot} \le  0, \quad  \dot{\varepsilon_{pl}} & = & 0
 .
 \label{h_scalaire}
\end{eqnarray}
The plasticity fraction, or  {\it yield function}, $h$, is zero at zero strain (elasticity only, no plasticity), and reaches 1 
 (plasticity only, no elasticity) at the yield strain,
which is a material dependent parameter.  
A smooth onset of plasticity can be described by a  plasticity function $h$ which continuously interpolates between the value 0 (completely elastic) and 1 (completely plastic)  \cite{pinceau}.

Eq. (\ref{h_scalaire}) closes the system of kinematic equations describing a foam. Indeed  from eqs (\ref{V_shared_scalar},\ref{h_scalaire}), we can write:
\begin{equation}
 {d\varepsilon_{el} }/{dt}  =  \dot{\varepsilon}_{tot} - h(  \varepsilon_{el} )  \dot{\varepsilon},
 \label{integrable}
\end{equation}
which can be integrated, knowing the applied strain rate, to calculate  the internal elastic strain.

The scalar constitutive equation   \cite{pinceau}
  proposed a total stress that is the sum of the elastic and viscous stress. 
 It was simple to assume a linear relation:
 \begin{equation}
\sigma=G \; \varepsilon_{el}+\eta \; \dot\varepsilon,
\label{sigma_pinceau}
\end{equation} with $G$ an elastic modulus and with $\eta$ a viscosity. 
 
 This model  is thus based on three material-dependent parameters; an elastic coefficient, a viscous one, and a plastic yield strain. It successfully predicted foam dynamical properties. 
 In oscillatory regime, it predicts  storage and loss moduli $G'$, $G''$. In stationnary regime, it resembles a Bingham model. It also enables to predict transient behaviours and creep \cite{pinceau}. 
 If needed, it is possible in principle to refine it, by introducing some non-linearities in eq. (\ref{sigma_pinceau}); for instance by  replacing   $\dot\varepsilon$ by  a power law, $\dot\varepsilon^n$
 (in which case, in a stationnary regime it looks like a Herschel-Buckley model); or
  by having $G$ or $\eta$ not constant.
  
\subsection{Direct measurement of strain, rearrangements and flow}
\label{direct_meas}

\begin{figure}[hbtp]
{\includegraphics[width=8.5cm]{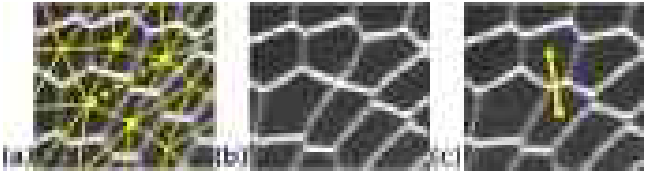}}
\caption{  Sequence of images of a  T1 event in a measurement box: (a) shows the network of links $\ellperso$ (thin lines) joining neighbour bubble centers  superimposed on snapshots extracted from a dry foam \cite{gra07}, and the disappearing link ($\ellperso_{d}$, dashes),  (b) the intermediate stage, and (c) the appearing link  ($\ellperso_{a}$, thick line).}
\label{fig:Topologicalchanges} 
\end{figure}

This section summarizes the parts of the companion paper \cite{gra07} which are used in what follows.
The matrices  $\deps_{tot}$, $\eps_{el}$, and $\deps_{pl}$ generalise the total, elastic and plastic strains in equation eq. (\ref{V_shared_scalar}), and are function of the position in the material. 
 They are measured directly using the statistical strains based on the bubble network:  $\tensor{U}$, $\tensor{V}$ and $\tensor{P}$, respectively. 
They  are all obtained from the texture tensor $\tensor{M}\equiv\langle \ellperso \otimes \ellperso \rangle$,  based on links $\ellperso$ between each pair of neighbouring bubble centers (see figure \ref{fig:Topologicalchanges}). Coarse-graining procedures yield a  continuous description of the foam. 
  
The internal strain  is defined from the relative deformation of bubbles with respect to the deformation at rest
$\tensor{M}_{0}$, as:
\begin{equation}
 \tensor{U}\equiv\frac{1}{2}\left(\log{\tensor{M}}-\log{\tensor{M}_{0}}\right). 
\label{icidefU}
\end{equation}


The topological rearrangment rate is defined from the {\em discontinuous} deformation of the network after each rearrangements. It describes the time derivative of $\tensor{U}$ due to T1 events: 
\begin{eqnarray}
\tensor{P}&\equiv& 
f\;\frac{1}{2}\left(\langle \ellperso_{d} \otimes  \ellperso_{d}\rangle -\langle \ellperso_{a} \otimes  \ellperso_{a}\rangle\right)\;\tensor{M}^{-1}+sym
\nonumber \\
&=&
f\delta\tensor{U}_\mathrm{T1}
,
\end{eqnarray}
$f$ being the frequency of rearrangments in an observation box containing one link, and $\delta\tensor{U}_\mathrm{T1}$  is the sudden drop in internal strain  associated with a T1. It is  computed from the disappearing links $\ellperso_{d}$ and appearing links $\ellperso_{a}$ (see figure \ref{fig:Topologicalchanges}).


The rate of strain is defined from the {\em continuous} deformation of the network.  It describes the time derivative of $\tensor{U}$ between T1 events as: 
\begin{eqnarray}
 \tensor{V}
 &\equiv&
 \frac{ \tensor{W}+ \tensor{W}^t}{2},\nonumber \\
{\rm where} \quad \tensor{W} &=& 
   \tensor{M}^{-1} 
   \left\langle   \vec{\ellperso}  \otimes \frac{d\vec{\ellperso}}{dt}  \right\rangle
 .
   \label{definiV}
\end{eqnarray}

\subsection{Experimental data}

Horizontal monolayers of bubbles (Fig. \ref{setups}) offer a unique possibility to observe all bubbles: their deformations, rearrangements and velocity. They are quasi-2D foams, because they are not strictly 2D. 
However, their flow is truly 2D: it has no vertical component.
A large set of detailed data is available; bubbles act as convenient tracers
of elastic strain, rearrangements and velocity
\cite{gra07,asi03,jan05}. 

We reanalyse experimental data already published  and courteously provided to us by the authors.
For details of the materials and methods, see the original publications. 
 In these four set-ups, coalescence
and ageing are below detection level. 
We assume that the gas and liquid constituents of the foam move together, at the same velocity (no drainage).

In  Fig. (\ref{setups}a), the foam is confined between the surface of water and a horizontal plate of glass. 
Bubbles are rather round, due to the high effective liquid fraction   \cite{Raufaste2007}. Thus  the region where T1s occur is larger, more widely distributed around the obstacle (compare Figs.   
(\ref{cap:plasticpredictedBenjamin})  and
(\ref{cap:plasticpredictedChristophe})  below).
There are thus more regions of the flow where statistics are significant.
This is why we use this experiment for the most detailed quantitative tests (section \ref{graphs}). 
In  Figs. (\ref{setups}b-d), the foam is confined between two horizontal plates of glass, and drier liquid fractions can be obtained. 

In Figs. (\ref{setups}a-c), the channel (only partly visible) is horizontal, its length is 1 m, its width 10 cm, its thickness 3.5, 3 and 0.5 mm respectively. It is filled with bubbles obtained by steadily blowing nitrogen in water with  commercial dishwashing liquid. 
 The bubbles are monodisperse and form a  disordered monolayer which reaches the free
exit at the end of the channel. The resulting steady plug flow, well in the quasistatic regime \cite{dol05PRE},
is made hetereogeneous by inserting a 3 cm diameter obstacle (Fig.
\ref{setups}a,b)
or a constriction  \cite{asi03}
(Fig.
\ref{setups}c). Thus different regions simultaneously display different
velocity gradients, elastic strains, and rearrangement rates, and allow to 
sample simultaneously many different conditions.

In Fig. (\ref{setups}d) the foam is in a 2D circular Couette geometry \cite{deb01}.
Briefly, the 2-mm thick foam monolayer is formed between two concentric disks
(only partly visible) with semicircular teeth of radius 1.2 mm to match the bubble diameter, thus anchoring bubbles at the walls. The outer disk, of radius 122 mm, is fixed. The inner disk, of radius 71 mm, rotates at 0.25 mm s$^{-1}$, thus shearing the foam. The resulting velocity field decreases quickly with the distance to the inner disk. This experiment has stirred  a debate about the existence and cause of velocity localization: for review, see for instance ref. \cite{jan06}. The experimental measurements we present here 
complement those of ref. \cite{jan05}; they are largely model-independent and might be used in the future to contribute to this debate.
  
\subsection{Identification with  continuous elastic, plastic and total strain rates}

This section discusses an additional property of foams, whether 2D or 3D: 
they are a unique material where the statistical measurements (section 
\ref{direct_meas}) can be identified with usual continuous quantities.

\subsubsection{Affine flow and total strain rate}
\label{affineGV}

\begin{figure}
(a)\\
 \includegraphics[width=7.2cm]{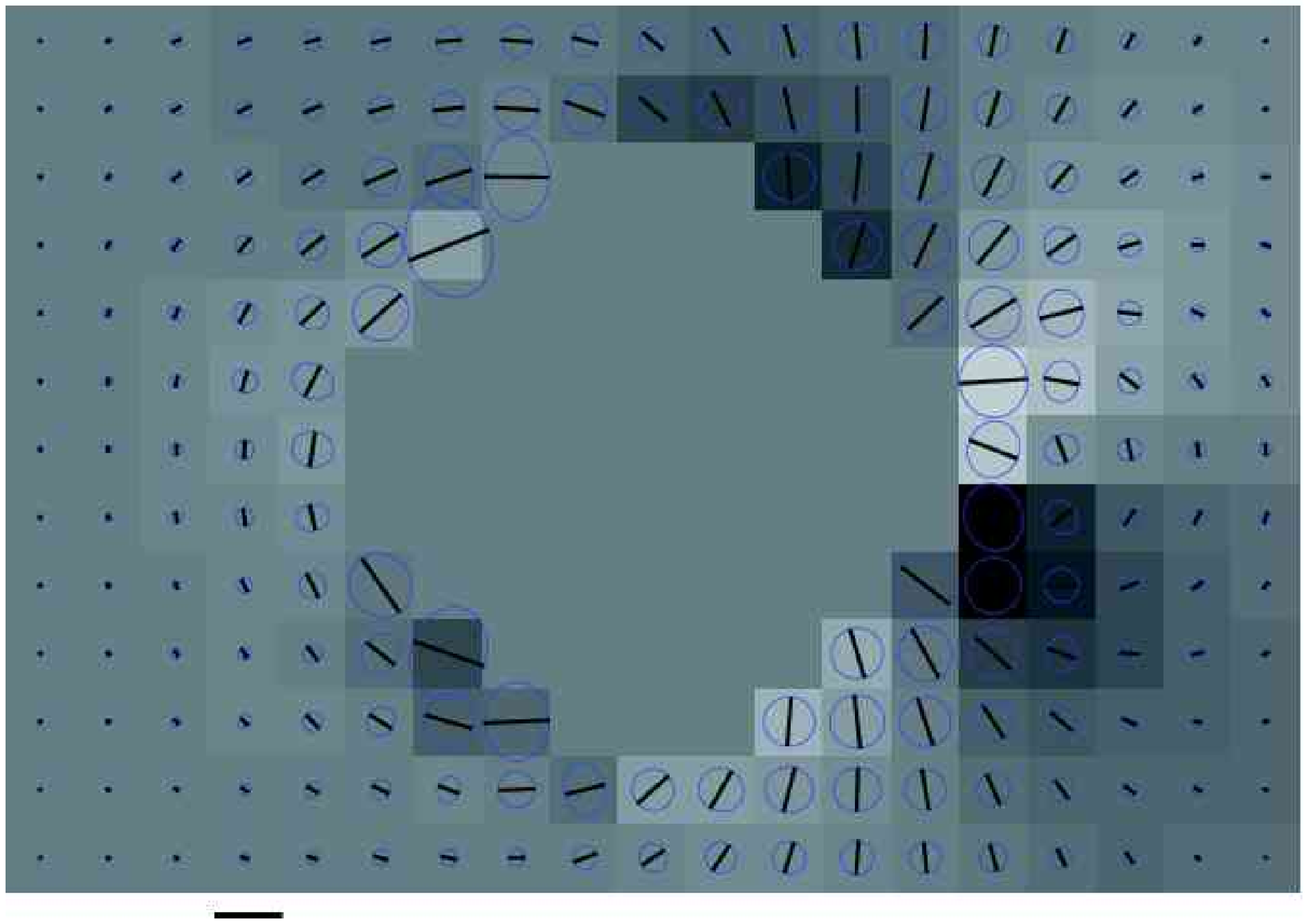}\\
 \includegraphics[width=8.8cm]{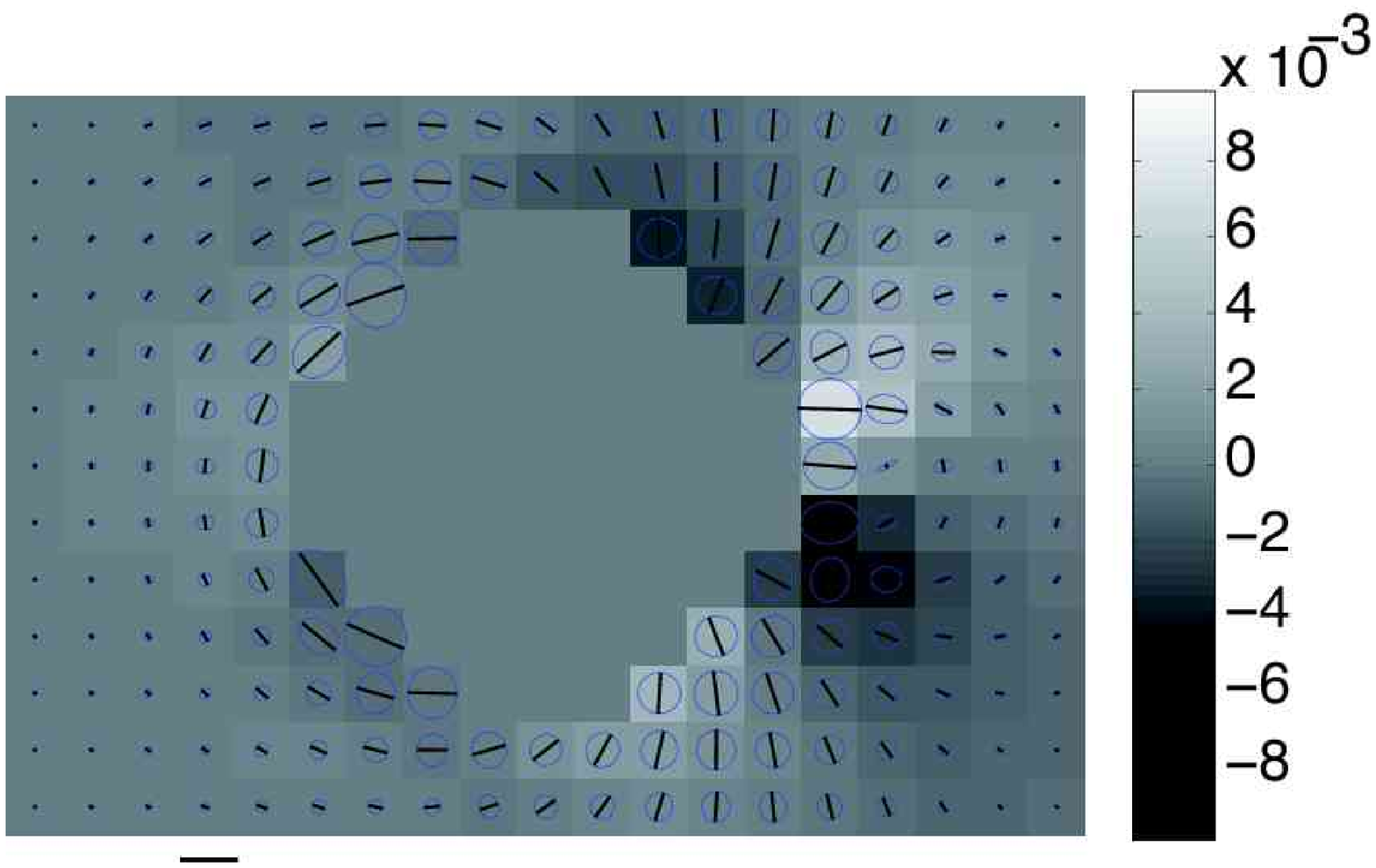}\\
(b)\\
 \includegraphics[width=6cm]{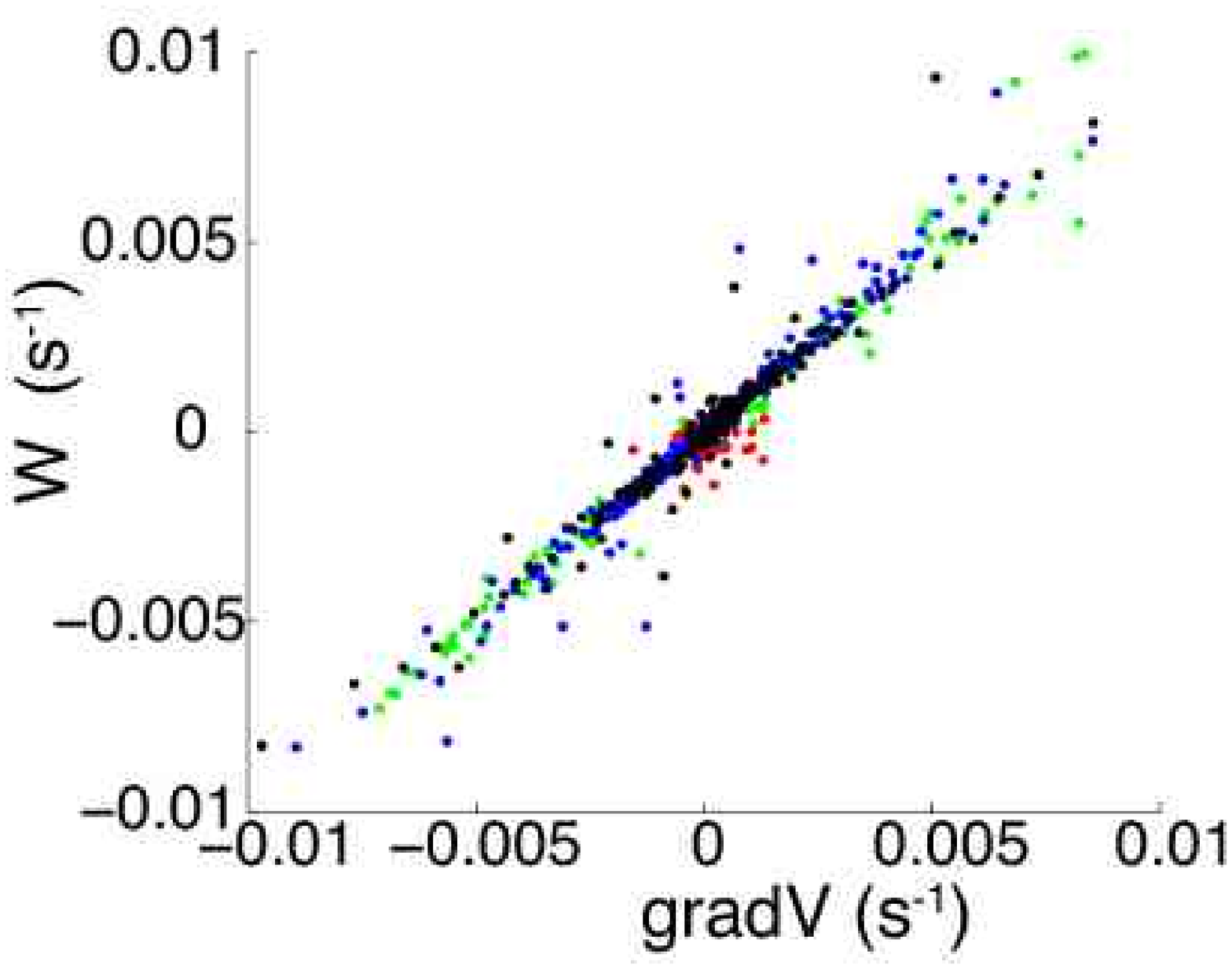}\\
 \includegraphics[width=6cm]{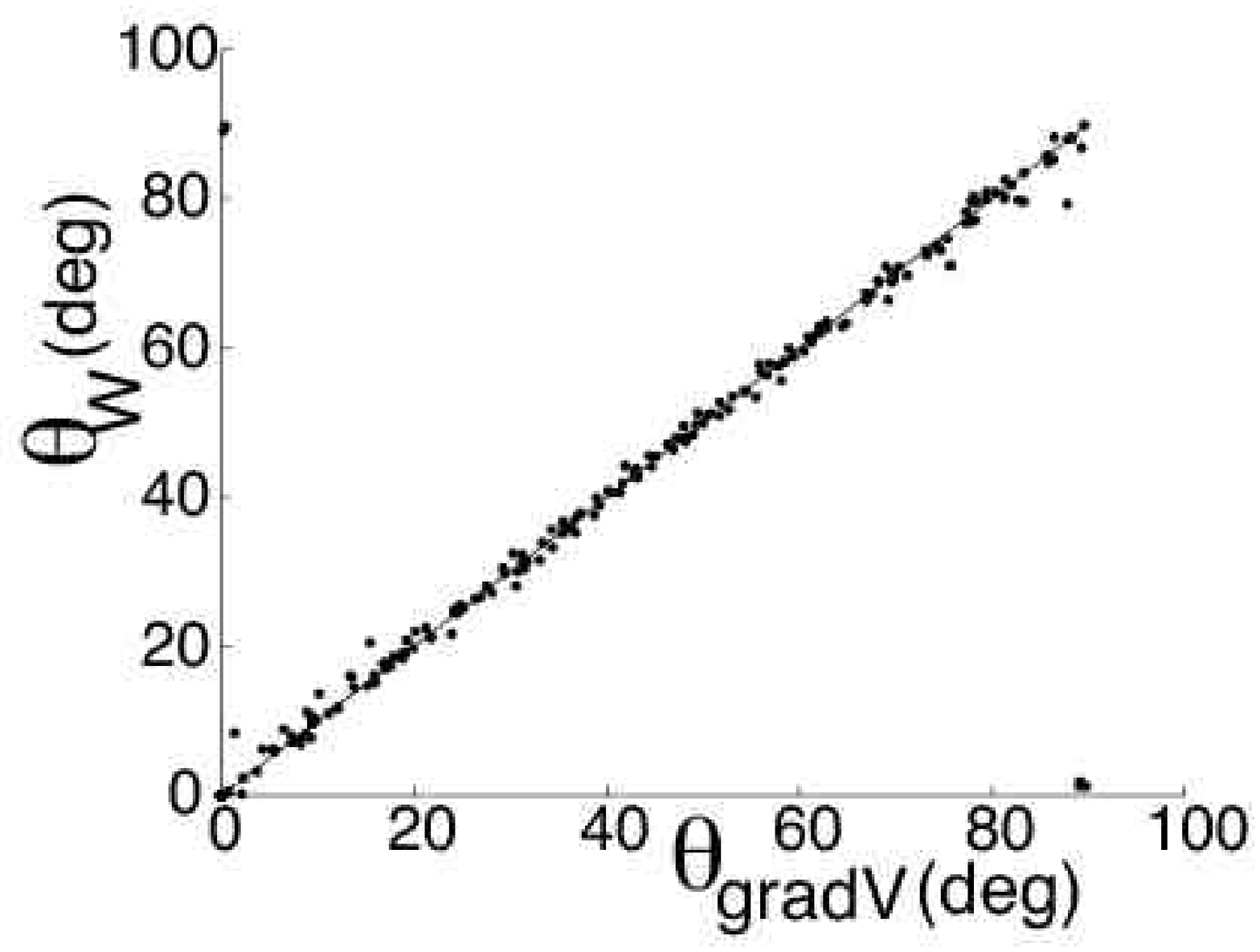}\\
\caption{ Affine assumption: 
test of eq. (\ref{G_affine}) on the  foam of Fig. (\ref{setups}b).
(a) Comparison of maps of   $   \tensor{W}$ (top) and $ \tensor{\nabla v}$ (bottom):
ellipses, symmetric part; grey levels, antisymmetric part (bar: $5.6\times 10^{-3}$ s$^{-1}$).
(b) Quantitative comparison: top, $(W_{XX} -W_{YY})/2$ {\it versus} $(\nabla v _{XX} -\nabla v _{YY})/2$, and 
the same for all other components (superimposed): $XX+YY$, $XY-YX$, $XY+YX$; 
bottom, angle (in degrees) of ellipses plotted in (a). Each point comes from one RVE of the image.
}
\label{gradVaffine}
\end{figure}

As discussed in the companion paper \cite{gra07}, if the flow of a material is affine, then $\tensor{W} $ measures the velocity gradient:
\begin{equation}
\tensor{W}   
\;    \stackrel{\rm affine}{\simeq}\; 
     \tensor{\nabla v}.
\label{G_affine}
\end{equation}
Fig.  (\ref{gradVaffine}) shows an example of our experimental tests of eq. (\ref{G_affine}).
We have measured the detailed components, including the rotational (asymmetric part), as well as eigenvalues and axes of the symmetric part. All these quantities are the same for $\tensor{W} $ and $ \tensor{\nabla v}$, within a few percents precision, with a  correlation close to 1.
The measurements of $\tensor{W} $ and $ \tensor{\nabla v}$ have a  comparable precision, and suffer from similar 
imprecisions near the channel walls and obstacle. Both have a small trace (ellipses are nearly circular).
Tests at smaller and larger scales, that is with different sizes of the
 representative volume element (RVE), 
 yielded similar results (data not shown). 

This agreement is unexpectedly good, given that
with the dry foam chosen here the strain is large and its gradient is strong. 
At large scale, movements of individual objects within the same RVE can differ considerably; but even in this unfavorable case, the affine assumption seems to hold, as shown in Fig.  (\ref{gradVaffine}).
The reason seems to be that eq. (\ref{G_affine}) is correct whenever $\tensor{M}$ does not vary significantly within the chosen RVE \cite{rau07}.
Together,  eqs.  (\ref{definiV},\ref{G_affine}) enable us to identify the rate of growth of links, $\tensor{V} $, with the total strain rate, $\deps_{tot}= (\tensor{\nabla v} +  \tensor{\nabla v}^t)/2$.

\subsubsection{Elastic stress  and strain}

\begin{figure}
 \includegraphics[width=8cm]{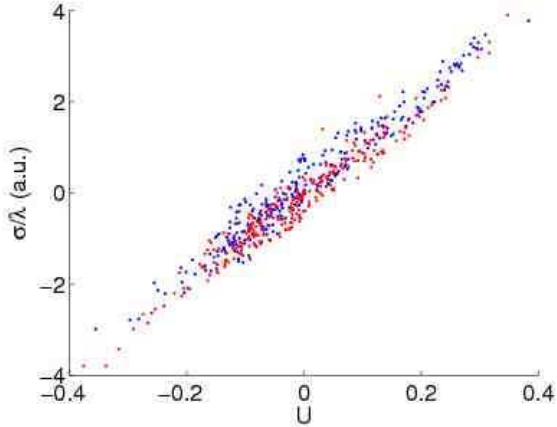}
\caption{ 
$XY$ components of the elastic stress $\sigma$ (in arbitrary units) {\it versus} $XY$ components of $\tensor{U}$. Each point comes from one RVE of the image of Fig. (\ref{setups}b).    The slope of the cloud of points is  the foam's 2D shear modulus; it is of order of the bubbles' line tension $\lambda$ to diameter ratio. Data for the $(XX-YY)/2$ components are superimposed: they have the same slope. }
\label{elastmodul}
\end{figure}
In a foam, the elastic energy is proportional to the bubble surfaces,
so that the elastic strain directly stems from bubble deformation.
It has been experimentally checked
that   $ \tensor{U}$ (or at least its deviatoric part)
   actually   determines the (deviatoric) elastic stress:
   see refs.  \cite{asi03,jan05}, to which we refer for  details of the measurement method. 
      We check it here too (Fig. \ref{elastmodul}), using our driest example (Fig. \ref{setups}b) in order to improve the measurement precision of the deviatoric elastic stress.

This indicates that the internal strain $\tensor{U}$, which is a state variable constructed from bubble deformations,  measures well the reversible strain $\eps_{el}$ that give rise to elastic stresses, although the flowing foam is clearly out of the elastic regime. In what follows, we thus call $\tensor{U}$ the "elastic strain".

Up to a prefactor, namely
the foam's shear modulus, $\scalarstrainsquared$ represents the elastic
energy stored ({\it e.g.} due to shear): that is, the difference between the energy of the current state and that of the local minimum. The latter varies at each T1.

\subsubsection{Plastic strain rate}

   \begin{figure}
  \includegraphics[width=7cm]{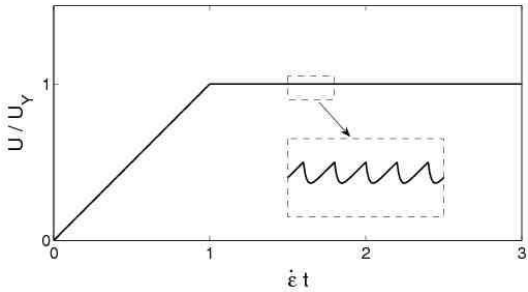}
  \caption{Schematic  impact of individual (discrete)
    rearrangements on the
stored elastic strain $U$, for a constant
loading rate $\dot{\varepsilon}$.
Rearrangements relax exponentially the
strain over a time $\tau_{\rm relax}$, 
with here $\dot{\varepsilon}\tau_{\rm relax}=0.02 \ll 1$.
In the present continuous model, rearrangements are coarse-grained. 
Reprinted from ref.  \cite{pinceau}.}
\label{fig:T1s}
  \end{figure}
   
 \begin{figure}[h]
 \setlength{\unitlength}{1cm} 
 \begin{picture}(8,7)(0.0,0.0)
\put(0,0){\includegraphics[width=7cm]{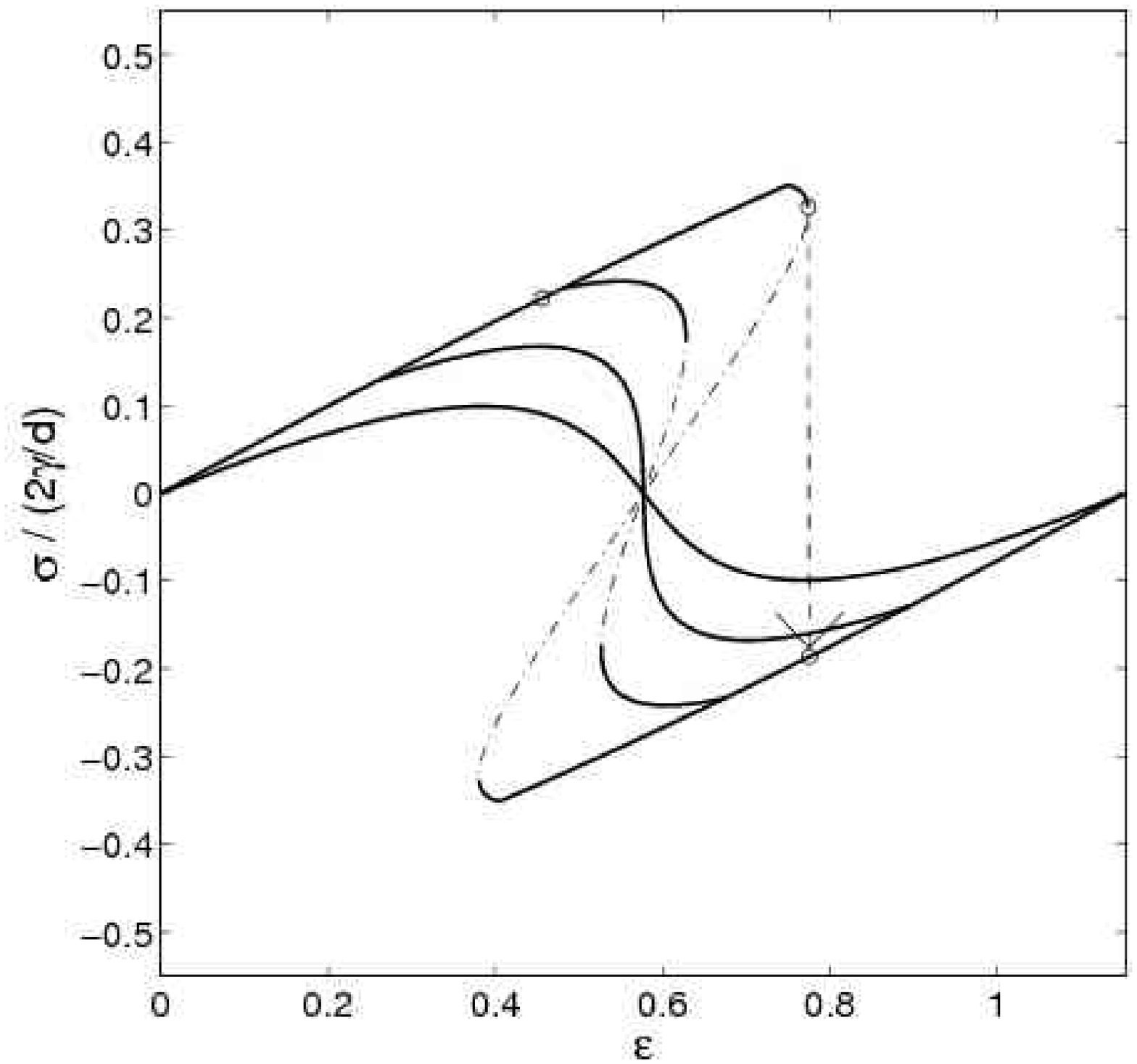}}
\put(2,5){\includegraphics[width=1.3cm]{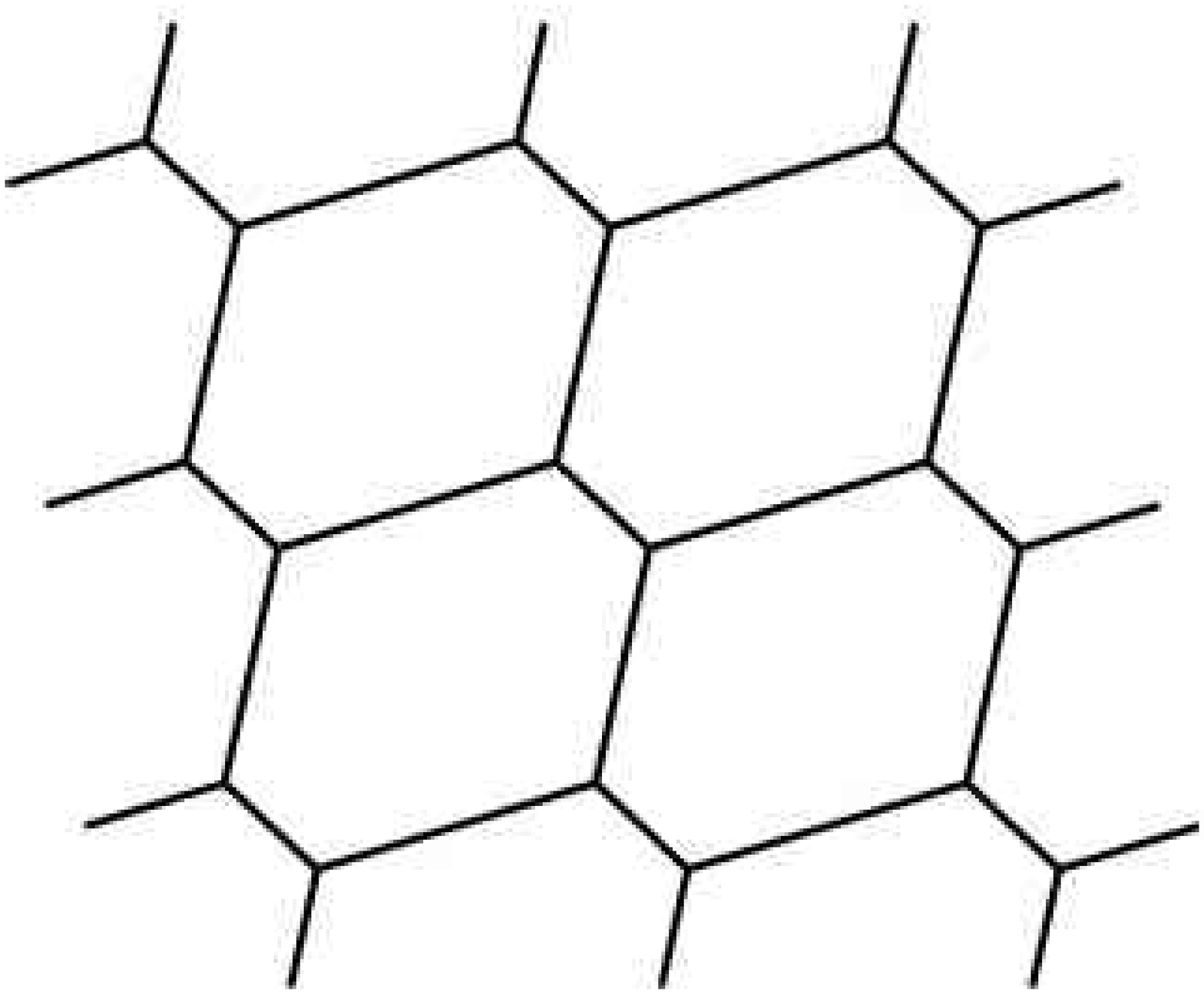}}
\put(4.5,5.6){\includegraphics[width=1.5cm]{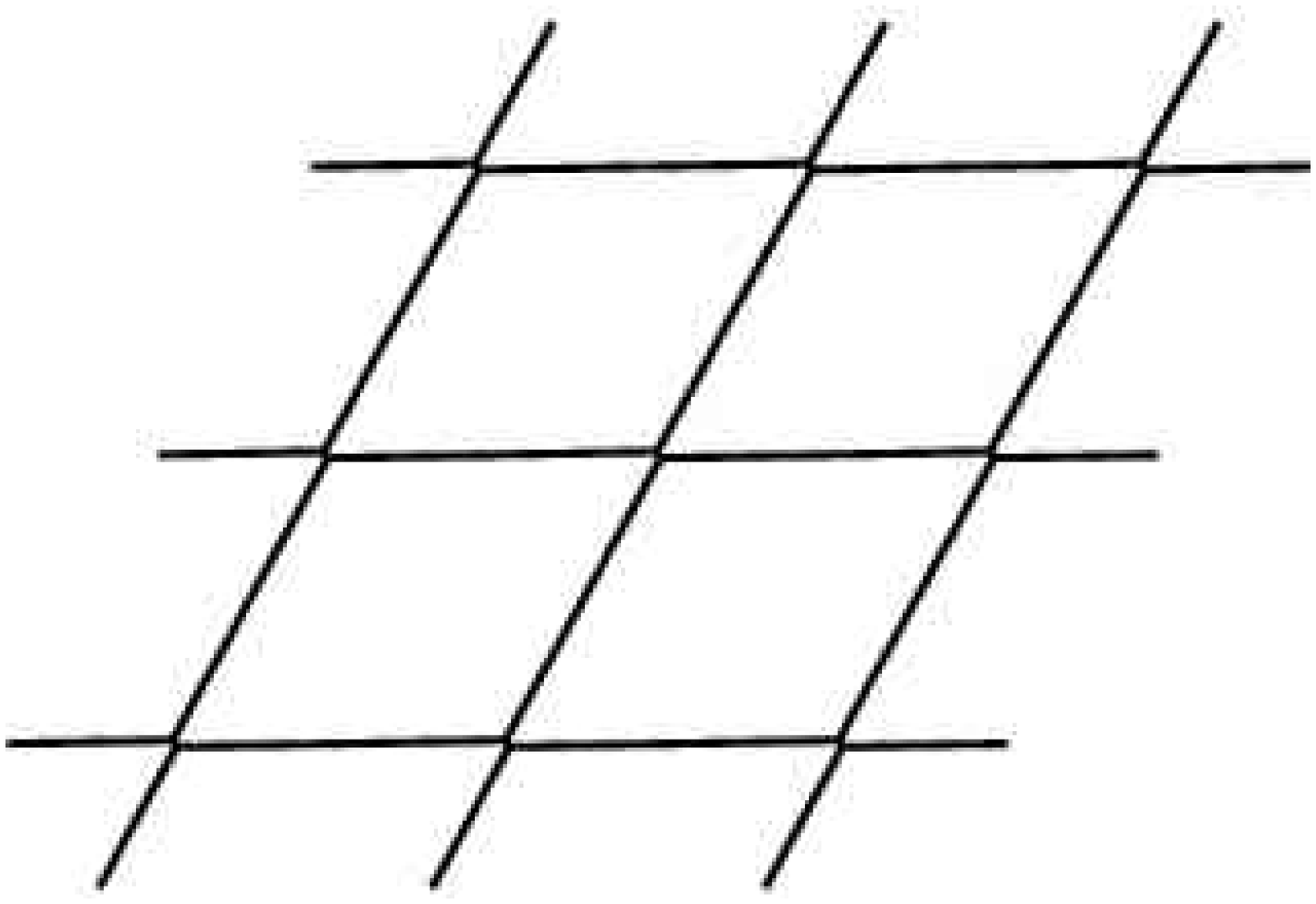}}
\put(5,1.5){\includegraphics[width=1.3cm]{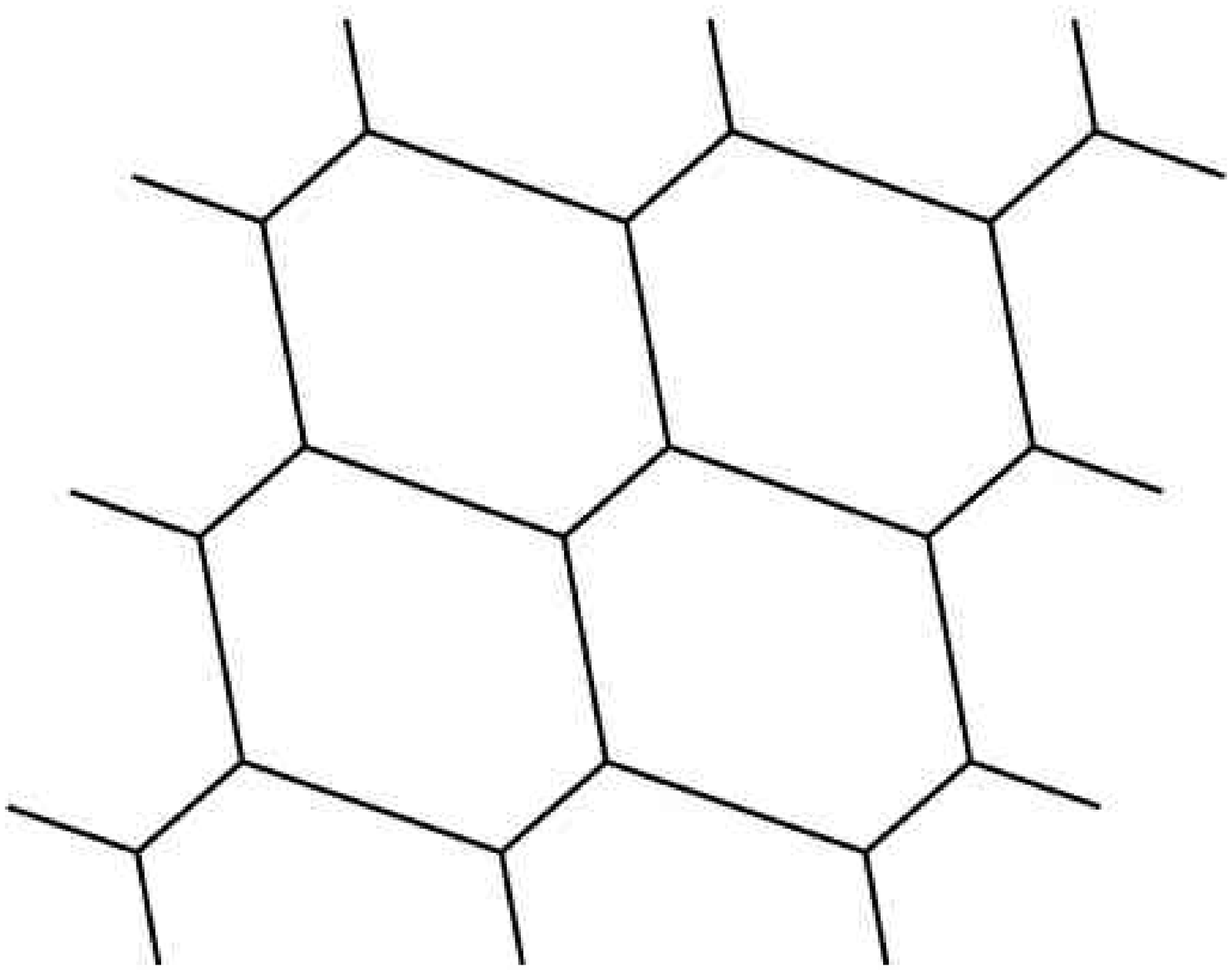}}
\end{picture}
\caption{Shear of  a 2D ordered foam \cite{prin83}, different liquid fraction: 9.3\% (close-packing, lowest amplitude curve), 5.5\% (onset of a negative slope unstable branch), 3\% and 1\%. Decreasing liquid leads to increasing maximum stress. The topological rearrangement occurs when the stress versus strain  curve becomes unstable (dotted line) and induces an irreversible strain. Three 1\% foams state are represented (circles on curve). The stress $\sigma$ is here normalised using surface tension $\gamma$ and bubble radius $d$.} 
\label{fig:Princen}
 \end{figure}
 
Disordered foams, which we consider here, are models for the plasticity of amorphous materials. 
Plastic events take the form of bubble rearrangements. When averaged over time or space, the effect  of topological rearrangements is usually smooth, and the foam behaves as a continuous material  (Fig. \ref{fig:T1s}). 
This contrasts with ordered foams, where bubbles are arranged in a honeycomb lattice and topological rearrangements are highly correlated, which are models for the plasticity of crystals based on dislocation movement \cite{Bragg1947,Gouldstone2001,BubbleRafts}. 

Each topological rearrangements is instantaneous (by definition of a topological change) and imply a change from a stable elastic branch to another. It is followed by a
 relaxation over a finite time $\tau_{\rm relax}$, determined by the ratio of dissipation  to driving elastic force.
  The rate of T1s is determined by the shear rate, which we keep here slower than  $\tau_{\rm relax}^{-1}$  in what follows (for extension to higher shear rates, see \cite{Saramito2007}).
 In that case, the foam has time to relax towards a new local equilibrium state  (Fig.  \ref{fig:Princen}), and the total energy dissipated is determined by the  difference between the energy barrier and the new local energy minimum. Thus the foam stays in this configuration: the transformation is plastic.
 It is possible to come back, with hysteresis, to the state  which preceded the T1: since it too is a local energy minimum, 
 the succession of a T1 and its inverse T1 leads to exactly the same pattern \cite{eli99,Lundberg2007}.
 
 Note that each relaxation following the T1 involves an {\em irreversible} dissipation, which measures the plastic dissipation rate (as seen on Fig.  \ref{fig:Princen}), if the strain rather than stress is imposed, the stress-strain curve of an 2D ordered foam with less that 5.5\% of liquid has an unstable branch). 
 The dissipation power of the plastic flow is proportional to the rate of T1s, thus to the shear rate:  the stress is thus independent on the shear rate, which is characteristic of solid friction (plastic contribution to stress, see plateau of  Fig. \ref{fig:T1s}).
 
 To summarize,  a topological rearrangement is equivalent to a plastic strain. This is specific to dry foams: as discussed in the  companion paper \cite{gra07}, this equivalence is not general to all materials.
  The irreversibility is associated with the relaxation after a rearrangement.




When a rearrangement occurs,  the total strain is not changed locally.
The elastic strain decreases by $\delta\eps_{el}=-\delta  \tensor{U}_\mathrm{T1}$,
and the plastic   strain thus  increases by
$\delta\eps_{pl}=\delta \tensor{U}_\mathrm{T1}$. 
Hence, the matrix $\tensor{P}$ measures well the plastic strain rate $\deps_{pl}=\delta \tensor{\varepsilon}_{pl}/\delta t$.

\subsubsection{Complete identification}

The preceding sections suggest that, in dry foams,  it should be possible to achieve the complete identification between statistical measurements and continuous quantities:
\begin{eqnarray}
\tensor{U}
&\; \stackrel{\rm foams}{\simeq}\; &
\eps_{el}
,\nonumber \\
\tensor{V}
&\simeq&
\deps_{tot}
,\nonumber \\
\tensor{P}
&\simeq&
\deps_{pl}
. 
\label{finalavectoutelatroupe}
\end{eqnarray}

The statistical measurements are linked by:
\begin{equation}
\tensor{V}
= 
\frac{{\cal D}{\tensor{U}}}{{\cal D}t}
+
\tensor{P},
\label{eq:dUdt}
\end{equation}
with ${\cal D}$ the total corotational derivative \cite{gra07}:
${\cal D}{\tensor{U}}/{{\cal D}t}$
$=$
${\partial\tensor{U}}/{\partial t}
+(\vec{v}\vec{\nabla})\tensor{U}
+\tensor{\Omega}\tensor{U}-\tensor{U}\tensor{\Omega}$.
Equation (\ref{eq:dUdt}) is the matrix version of eq. (\ref{V_shared_scalar}); here too, the total applied strain rate is the sum of the internal elastic strain rate and the irreversible plastic strain rate.

 We conclude that  the measurable quantities  
 $\tensor{U}$, $\tensor{V}$ and $\tensor{P}$ 
 make the connection between the discrete pattern 
and its continuous mechanical behaviour. In the following,
since we present experimental results, 
we stick to the statistical measurements of strains (l.h.s. of eqs. (\ref{finalavectoutelatroupe})).
They can of course be replaced by the usual notations 
from material science (r.h.s. of eqs. (\ref{finalavectoutelatroupe})).

\section{Model for foam plasticity}
\label{plasticmodel}

This section introduces a prediction of the plastic strain rate and topological rearrangements in foams. 

\subsection{Plastic strain rate}

The deviatoric elastic strain (shear, without dilation) is defined as the traceless
matrix:
\begin{equation}
\tensorstrain_{d}\equiv\tensorstrain-\frac{1}{2}({\rm Tr}\tensorstrain)\tensor{I}.
\end{equation}
Its amplitude is defined as 
\begin{equation}
\scalarstrain\equiv\left[\frac{\tensorstrain_{d}:\tensorstrain_{d}}{2} \right]^{1/2},
\end{equation}
where we use the double contraction product, namely the scalar product of matrices, $\tensor{A}:\tensor{B}= \sum_{i,j}{A}_{ij}{B}_{ij}.$ In a 2D configuration, $U_{d}$ provides the absolute value of the eigenvalues of matrix $\tensor{U}_{d}$.
The matrix $\tensorstrain_{d}/\scalarstrain$ is then  a directional 
matrix that writes $diag(1,-1)$ in the eigenvector basis of elongation. For details of notations see \cite{gra07}.

To generalize the scalar model, we need to specify not only the amplitude of the plastic strain rate matrix $ \tensor{P}$, which we take linear in the strain rate  $\tensor{V}$; but also its 
 direction, which we assume is  aligned with the current deviatoric elastic strain $\tensor{U}_{d}$.
 In other words, the plastic evolution is directed along the preexistent elastic strain and occurs with a rate which is the projection  of the total strain rate onto the elastic strain.
As will become apparent below (section \ref{mean field}), this amounts to a  mean field approximation. 

  Our main assumption  is thus that $ \tensor{P}$ is determined by the "projection" of  $\tensor{V}$ on $\tensor{U}_{d}$, defined (in analogy with the projection of a vector on another) using the double contraction product:
 \begin{equation}
\tensor{V}_{proj}=\frac{({\tensor{V} : \tensorstrain_{d}})
}{2{U_{d}}^2}  \:\tensorstrain_{d}.
\label{eq:Vproj}
\end{equation} 
 We thus generalise eq. (\ref{h_scalaire}) using  a scalar  plasticity function, $h$,  of the strain amplitude,  $U_{d}$:
\begin{equation}
\mathrm{if} \quad \tensor{V} : \tensorstrain_{d} > 0, \quad \tensor P=  
{h(U_{d})}\tensor{V}_{proj}
 ,
\label{eq:Ppredicted} 
\end{equation}

Eq. (\ref{eq:Ppredicted})   applies for the case where the total strain loads the internal strain ($\tensor{V}: \tensor{U}_{d} $ is positive). 
If $\tensor{V}$ is proportional  
to $\tensor{U}_{d}$ (same direction, same anisotropy), this equation reduces to  $\tensor{P}=h(U_{d})\tensor{V}$,  equivalent to the scalar one, eq. (\ref{h_scalaire}); moreover, if  $U_{d}$ has reached the yield value ($h=1$), then $\tensor{P}=\tensor{V}$.
 
On the opposite,   if the
applied strain rate $\tensor{V}$ is in the direction opposed to internal strain ($\tensor{V}:\tensor{U}_{d} $ is negative), it  contributes
to unload it elastically.
 It  thus does not induce many  rearrangements   \cite{rau07}. We neglect them by setting  $\tensor P$ to zero :
 \begin{equation}
\mathrm{if} \quad \tensor{V} : \tensorstrain_{d} < 0,  \quad \quad \tensor P= 0.
\label{P_partI}
\end{equation}

 Note also that eq. (\ref{eq:Ppredicted}) 
would reduce to the classical Prandtl-Reuss model for  a perfect plastic material that yields when the elastic strain reaches the value $U_Y$  \cite{was75}, 
if the plasticity function   was an Heaviside function, $h\left(U_{d}\right) = {\cal H}(U_{d}-U_Y)$,
discontinuously jumping from the value 0 when $U_{d}<U_Y$ to 1 when $U_{d}\geq U_Y$.


\subsection{Rearrangement frequency}
   
 Each rearrangement modifies the strain in its measurement box.
We consider here that  the strain is  concentrated within the reference area attributed to one link $A_{link}=A/N_{link}$ (the perturbation in a continuous elastic space rapidly decays with the distance to the rearrangement location, see model \cite{pic04}). If we assume that each plastic change has a constant amplitude $\varepsilon_{0}$, we can write:
\begin{equation}
 \delta  \tensor{U}_\mathrm{T1}=\varepsilon_{0}\; \frac{\tensor{U}_{d}}{U_{d}}, 
 \label{hypdeltaU}
 \end{equation}
where we assumed that rearrangements are aligned with elasticity (following eq. \ref{eq:Ppredicted}).
We have seen that the plasticity rate  matrix can be written:
 \begin{equation}
 \tensor{P}=f\; \delta \tensor{U}_\mathrm{T1}, \label{eq:PfdeltaU}
\end{equation}
where $f$ is the frequency of rearrangements {\em per link}. When considering averages in  larger counting boxes,    containing $N_{link}$ links (approximately 3 times the number of bubbles in the box \cite{wea99}), equation (\ref{eq:PfdeltaU}) still holds and writes $ \tensor{P}=f_{box}\; \delta \tensor{U}_{box}$. Indeed  the frequency in the counting box is $f_{box}=N_{link}f$, and  the impact of a T1  on a larger surface is diluted to the value $ \delta  \tensor{U}_{box}=\delta  \tensor{U}_\mathrm{T1}/N_{link}$. 

 
Combining the plasticity equation (eq. \ref{eq:Ppredicted}) and the amplitude of strain relaxation (eq. \ref{hypdeltaU}), we obtain the frequency $f$ of T1 events, per link:
\begin{eqnarray}
\mathrm{if} \quad \tensor{V}:\tensor{U}_{d} > 0, \quad
f &=& \frac{h(U_{d})}{2\varepsilon_{0}{U_{d}}}  \left( \tensor{V}:\tensor{U}_{d} \right)
,\nonumber \\
 \mathrm{if} \quad \tensor{V} : \tensor{U}_{d} < 0, \quad  f & = & 0
 .
 \label{eq:fT1}
\end{eqnarray}
 It depends on the positive eigenvalue  $\dot{\gamma}$ of the elongation
rate, and on the relative angle $\theta$ between the eigenvectors of the elongation
rate and strain.
If  $\cos(2\theta) < 0$, the frequency is zero; 
if $\cos(2\theta) > 0$, the order of magnitude of the frequency can be estimated:
 $ f\propto\cos(2\theta)\dot{\gamma}$.
This extends findings by \cite{vin06}. It expresses that rearrangements
are frequent where the total strain rate is strong, and when the elongation rate
is parallel to the pre-existing strain thus loading it through
the yield surface.


\section{Experimental tests}
\label{tests}

\subsection{Rearrangements: orientation and frequency}


This section uses data of  Fig. (\ref{setups}a).

\label{graphs}

\begin{figure}[htbp]
{\includegraphics[width=7cm]{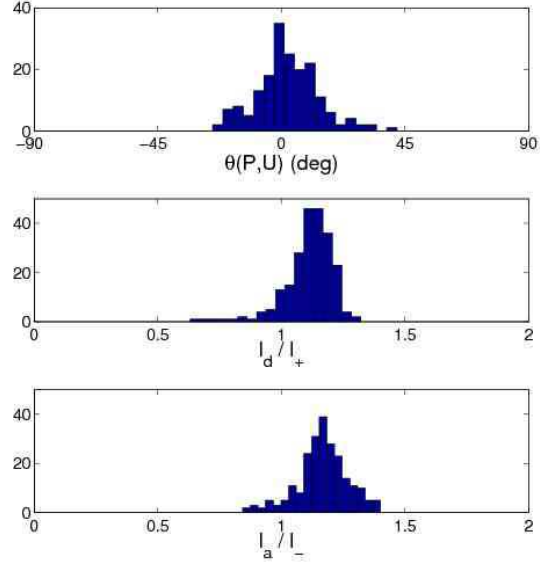}}
\caption{Histograms of measurements
in all regions of the foam (Fig. \ref{setups}a). Top: angle between topological events
and elongation. Middle: length of disappearing links compared to average
length in elongation direction $\ell_{+}$. Bottom: length of appearing
links compared to average length in compressed direction $\ell_{-}$.}
\label{fig:figure17anglelpluslmoinsVer}
\end{figure}

Fig. (\ref{fig:figure17anglelpluslmoinsVer})   confirms
the mean field approximation of the model (eq. \ref{eq:Ppredicted}), 
namely 
that disappearing and appearing links are determined (in average)
by the existing strain. More precisely,  
$\tensor{P}$ makes an angle of $0\pm9^{\circ}$ with $\tensor{M}$
or $\tensor{U}$: disappearing links $\vec{\ell}_{d}$ are mainly
in the elongation direction (Fig. \ref{fig:figure17anglelpluslmoinsVer}, top). Their length is 1.2$\pm$0.1 times larger
than the average of existing links in that direction, $\ell_{+}= \sqrt{2\lambda_1}$, where $\lambda_1$ is $\tensor{M}$'s largest eigenvalue (Fig. \ref{fig:figure17anglelpluslmoinsVer}, middle). Conversely,
links $\vec{\ell}_{a}$ appear in the contracted direction of $\tensor{M}$, with a
length 1.1$\pm$0.1 times the average of existing links, $\ell_{-}= \sqrt{2\lambda_2}$ (Fig. \ref{fig:figure17anglelpluslmoinsVer}, bottom).
\label{mean field} 

The jump in elastic strain is therefore  oriented along $\tensor{U}$. Its amplitude is approximately:
\begin{eqnarray}
  \delta \tensor{U}_\mathrm{T1} 
&\approx&
\left(
\begin{array}{cc}
  \ell_{d}^2 / \ell_{+}^2
 &0\\
0&
 -\ell_{a}^2/\ell_{-}^2
  \end{array}
\right)
\nonumber\\
&\approx&
\left(
\begin{array}{cc}
1.3\pm0.2 &0\\
0&-1.2\pm0.2
  \end{array}
\right),
\label{approx unit strain}
\end{eqnarray}
using  $ \delta \tensor{U}_\mathrm{T1}\approx\delta \tensor{M}\times \tensor{M}^{-1}/2$, from the differentiation of eq.
(\ref{icidefU}) \cite{gra07}.
  We conclude from equation (\ref{hypdeltaU}) that each rearrangement changes the elastic strain per link
by a constant amount:
\begin{equation}
\varepsilon_{0}\simeq 1.2 \pm 0.2.
\label{unit_strain}
\end{equation}
Strain is decreased by slightly more than one average
length in the elongation direction, and increased by slightly more than one average length
in the orthogonal direction.

\begin{figure}[htbp]
\includegraphics[scale=0.5]{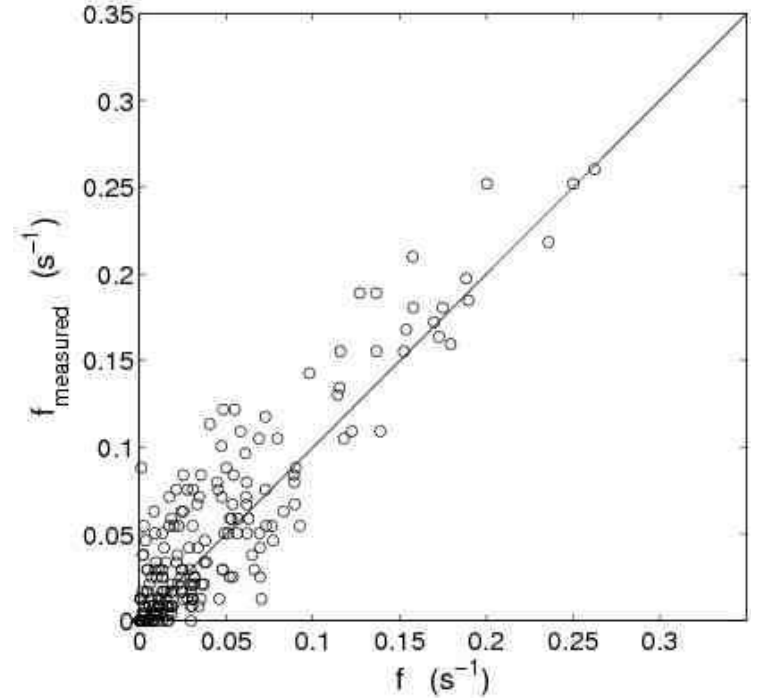} 
\caption{ Frequency of rearrangements:  
 observed  \textit{versus} predicted.
 Each point corresponds to a RVE of the foam, that is, one ellipse
of  Fig. (\ref{cap:plasticpredictedBenjamin}). 
Observations:  frequency $f_\mathrm{measured}$ of rearrangements (per link and second) mesured on Fig. (\ref{setups}a).
Predictions: $f$ from 
eq. (\ref{eq:fT1}), setting the yield function to $h=\min((U/U_Y)^2,1)$ and the yield $U_Y=0.15$, while  $\varepsilon_{0}=1.2$ (eq. \ref{unit_strain}).
Solid line: diagonal $f_\mathrm{measured}=f$. 
\label{cap:tests} }
\end{figure}

Rearrangement frequency is well predicted (Fig.
\ref{cap:tests}) by equation (\ref{eq:fT1}).
 The main parameter required, 
namely $U_Y$, is directly read from measurements of $U$: here $U_Y=0.15$, which is reasonable for a foam with 4\% liquid fraction. Second, the shape of the elasto-plastic transition has been chosen as a quadratic $h$ function, which is justified in section
 \ref{maps}.

%

 
\begin{figure}[htbp]
\hspace{0.4cm}\includegraphics[width=6cm]{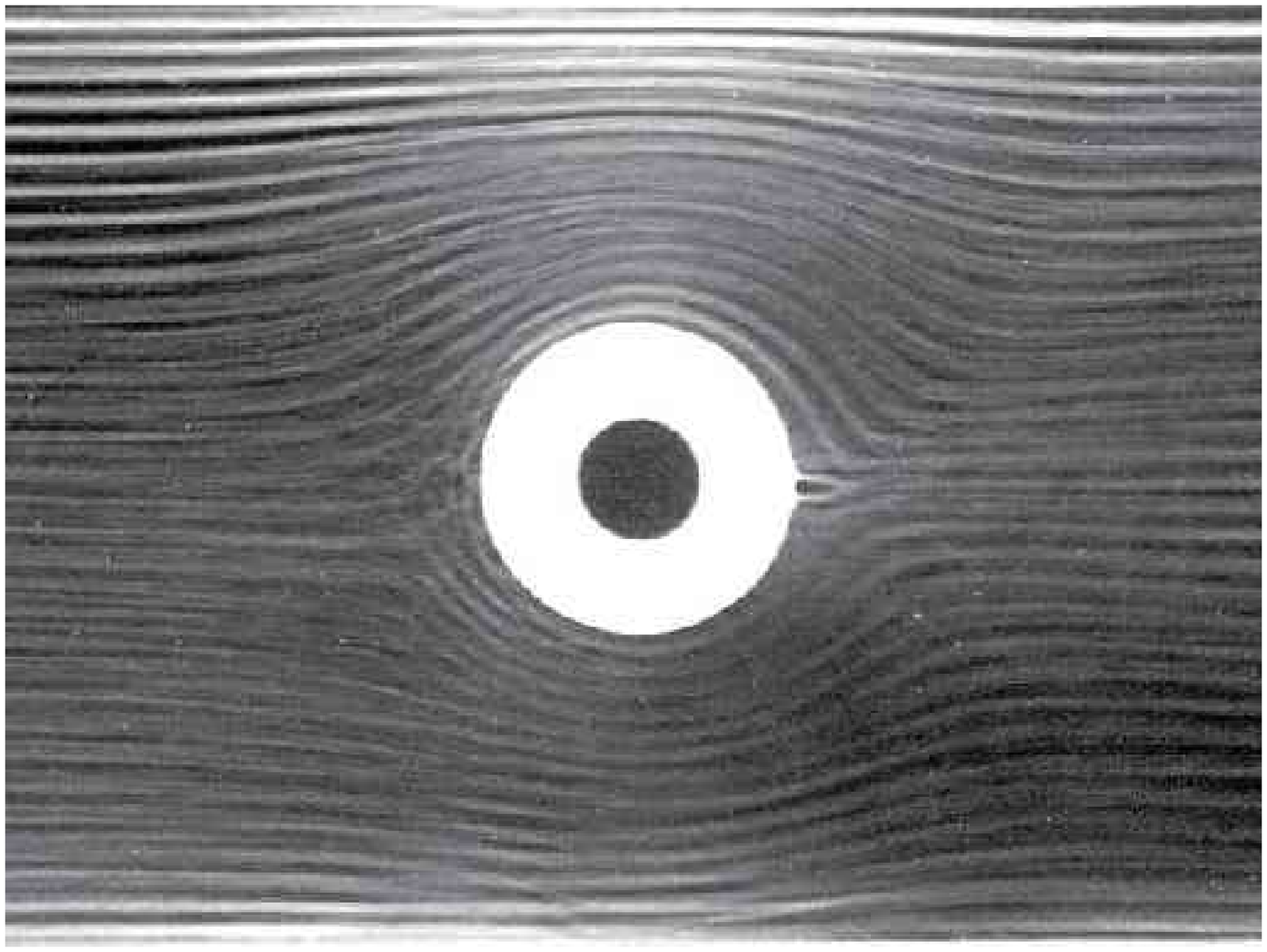} \\
 \includegraphics[width=6.5cm]{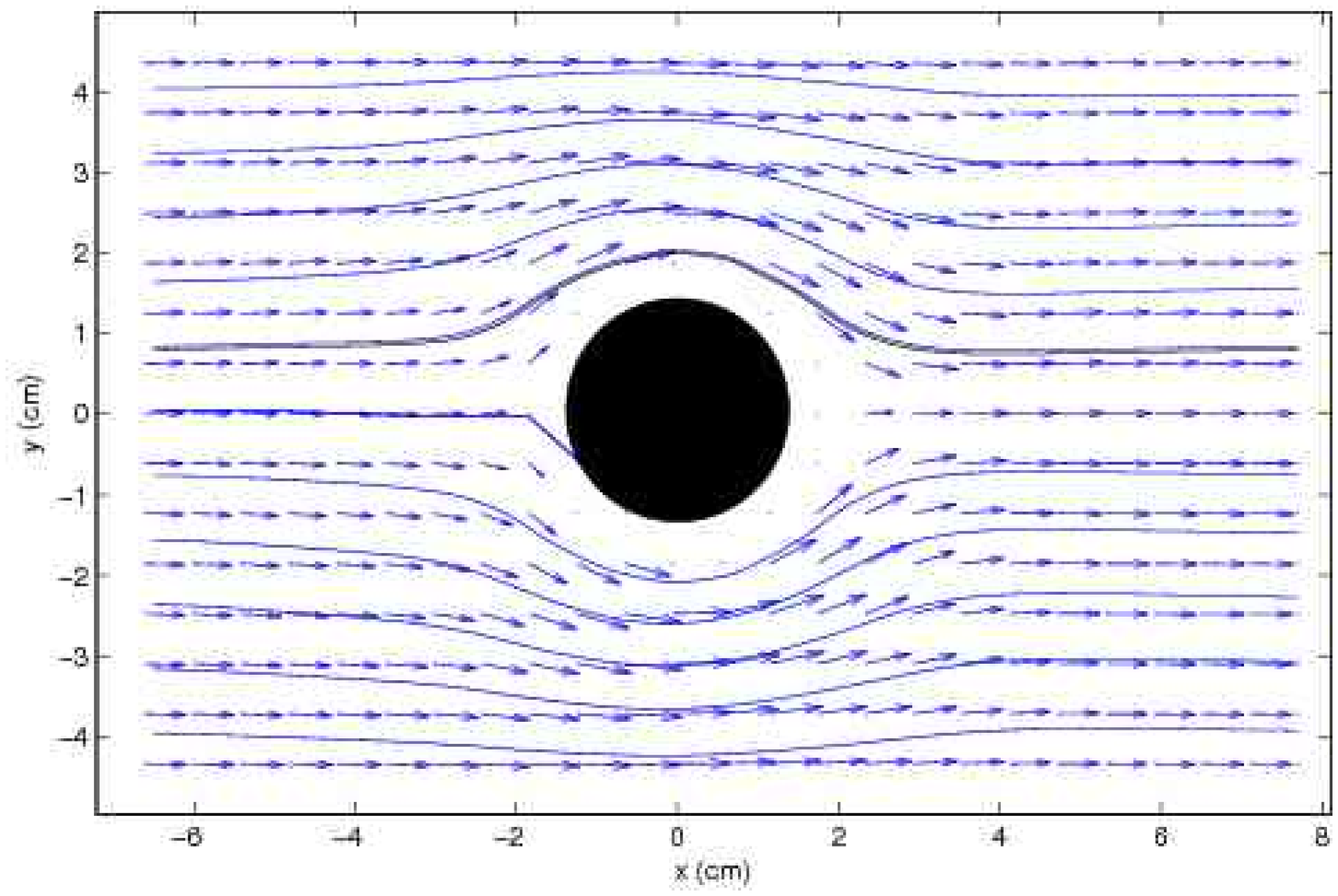} 
\caption{ Foam flow. Top: superimposed images evidence the streamlines (here from Fig. (\ref{setups}b)).
Bottom: measured velocity field and streamlines. The solid line highlights the streamline analysed in Fig.  (\ref{topo_streamline}).}
\label{streamline}
\end{figure}
 
\begin{figure}[htbp]
 \includegraphics[width=8cm]{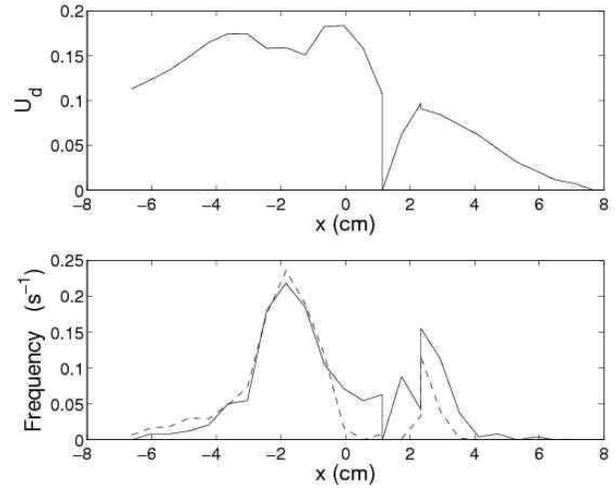} 
\caption{Measurements along the streamline shown on Fig. (\ref{streamline})bottom. Top: loading and unloading of elasticity, $U_d$. Bottom: T1 frequency along streamline, as in Fig. (\ref{cap:tests}); solid line: experimental $f_\mathrm{measured}$; dashes,
predicted $f$.}
\label{topo_streamline}
\end{figure}

The origin of the upstream/downstream asymmetry of plasticity can
now be qualitatively explained, by following a bubble along its streamline (Fig. \ref{streamline}). The obstacle imposes a succession of opposite
elongation rates: spanwise before the obstacle, and streamwise after
it (see below, Fig. \ref{cap:plasticpredictedBenjamin}). 
Before the obstacle, the elastic strain
and the elongation rate are aligned, and
the foam is yielding.
 After the obstacle,
it takes some time until the elastic strain fully relaxes, then increases again
in the new direction of elongation, orthogonal to the initial one (Fig. \ref{topo_streamline}).
Topological rearrangements are therefore concentrated in a smaller
region. 

\subsection{Maps}
\label{maps}

\begin{figure}[htbp]
\begin{center}
\includegraphics[width=7.8cm]{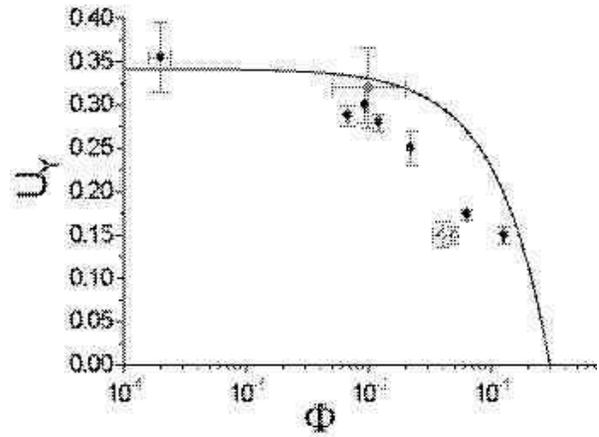}
\caption{Yield strain versus liquid fractions for the different setups of Fig. \ref{setups}: $\circ$, (a); $\bullet$, different experiments of (b); $\square$, (c); $\vartriangle$, (d). Trend of  data (b) can be qualitatively  compared to a quadratic law \cite{Mason1996}: $U_Y = C (\Phi_{c}-\Phi)^2$, with $\Phi_{c}$ corresponding to the rigidity limit liquid fraction ($\approx$ 30 \% for setup (b) \cite{gra07}) and 
$C$ an empirical constant equals to 1.1.}
\label{fig:UyPhi-cartes}
\end{center}
\end{figure}

This section presents the spatial distribution of measurements plotted as ellipse maps (matrix fields), which simultaneously display: position, orientation, anisotropy and frequency of rearrangements.
Again, we   predict plasticity from the measured elastic strain and total strain rate, using a yield strain $U_Y$ directly read from the onset of a plateau in the plasticity fraction $h$, without adjustable parameter (see Fig. \ref{fig:UyPhi-cartes} for the different measurements). A smooth quadratic plasticity fraction $h$ is chosen (as observed from rheometric measurements \cite{pinceau}), to account for possible plastic events below the yield.
This {\it direct} prediction of  $\tensor{P}$ from  $h$ is robust: in fact, it is especially significant near the yield strain, where $h$ is close to 1 (whatever the choice of $h$).

Conversely, the {\it inverse} estimate of $h$, from $\tensor{P}$, is noisier. In fact, differences between various candidates for the $h$ function are more significant far from the yield point, where measurements are both smaller and with less statistics.
In practice, we suggest to estimate the amplitude of deviatoric part $\tensor{P}_{d}$ of plastic strain with the amplitude $P_{d}=(\tensor{P}_{d}:\tensor{P}_{d}/2)^{1/2}$; and the amplitude of the deviatoric part of the total strain rate with $V_{d}=(\tensor{V}_{d}:
\tensor{V}_{d}/2)^{1/2}$.
If we assume  $\tensor{P}$ and $\tensor{V}$ are nearly parallel (which is not really the case with obstacles), 
eq. (\ref{eq:Ppredicted}) can be projected on the same axis:
we obtain an estimate of the plasticity fraction as  $h\simeq P_{d}/V_{d}$.


\subsubsection{Flow around an obstacle: wet foam}
\label{flow_wet_foam}

\begin{figure}[htbp]
(a)
 \includegraphics[width=6.5cm]{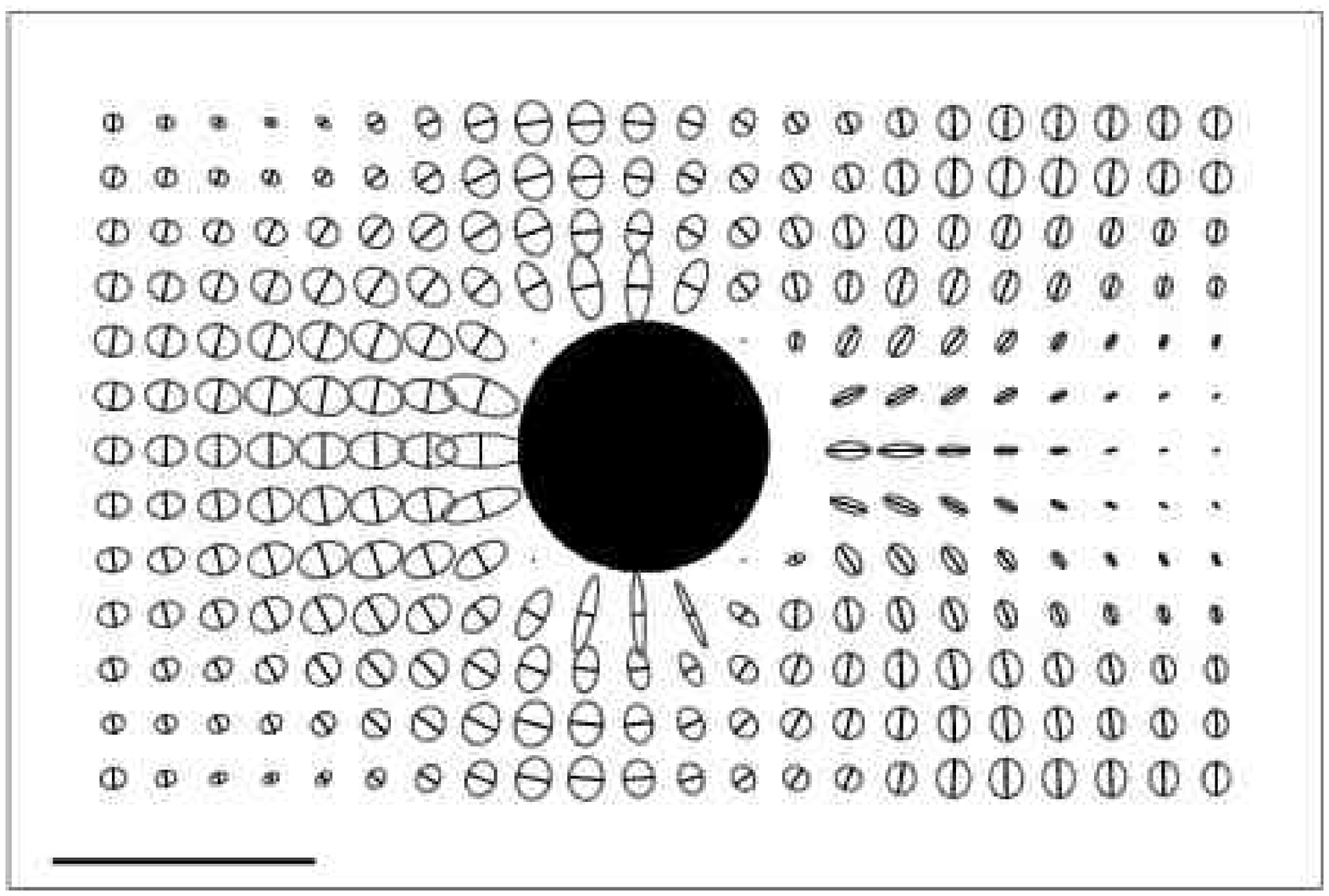}\\
 (b)
\includegraphics[width=6.5cm]{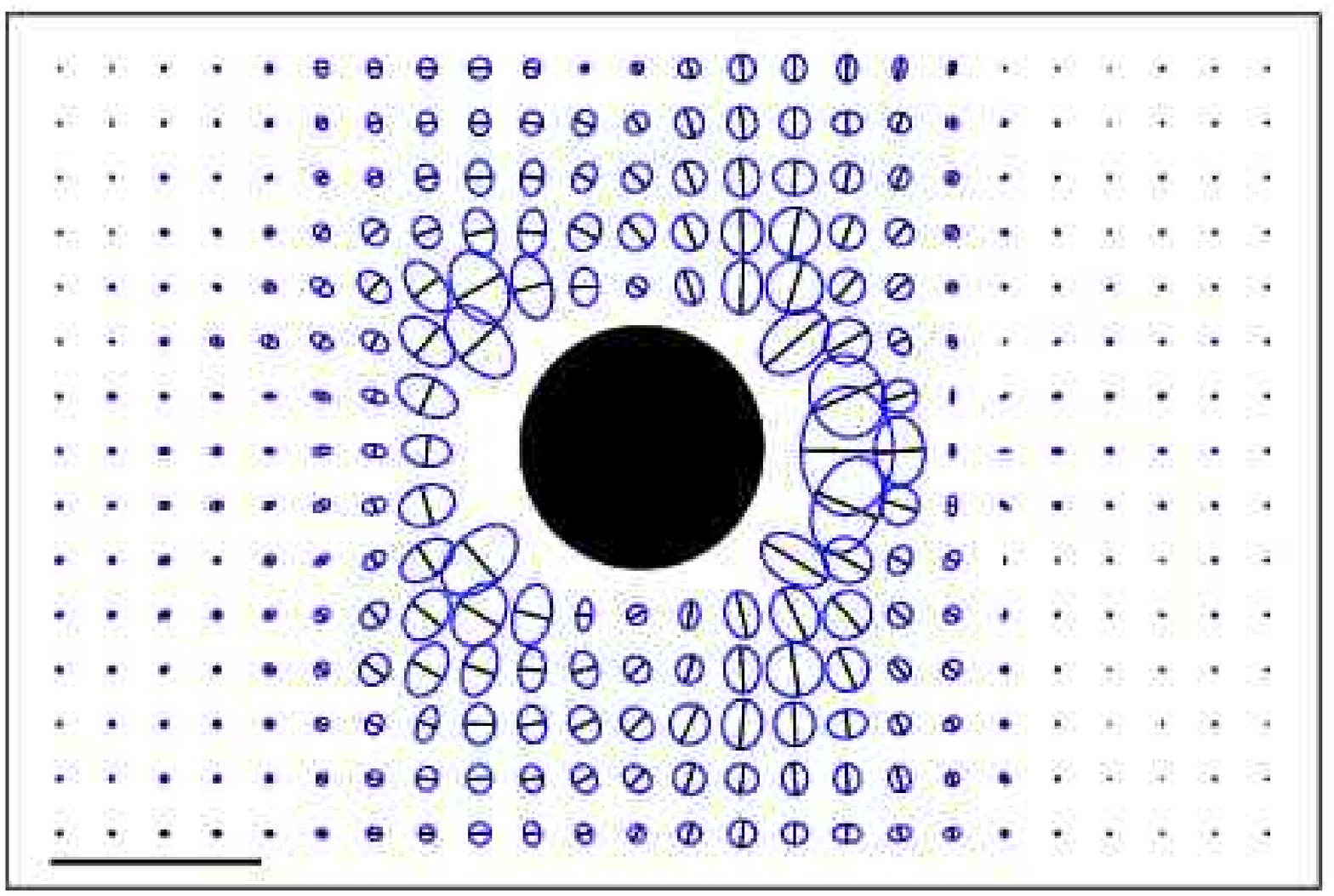}\\
(c)
\includegraphics[width=6.5cm]{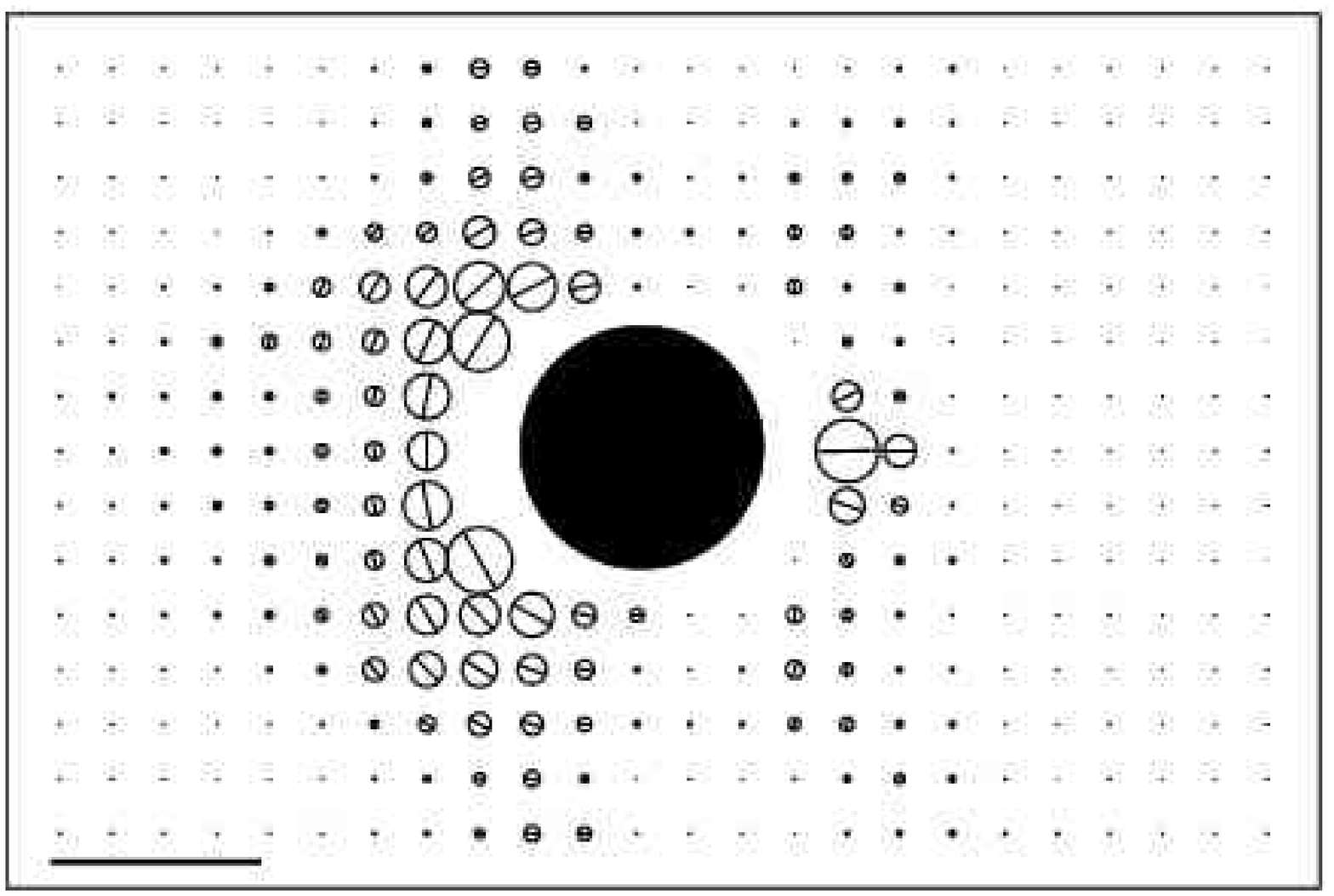}\\
(d)
 \includegraphics[width=6.5cm]{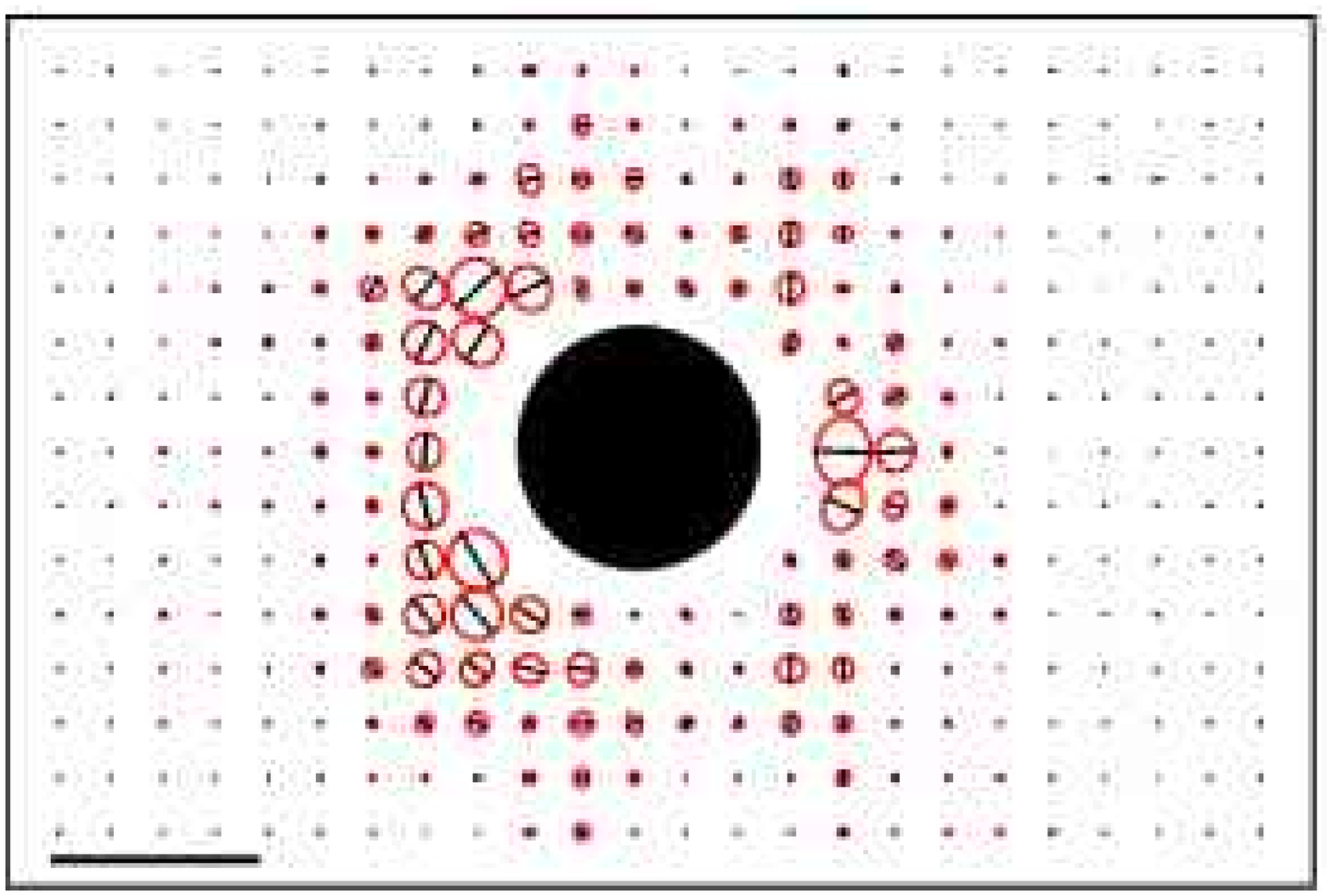}\\
\caption{Wet foam flowing around an obstacle (Fig. \ref{setups}a). Measurements of  $\tensor{U}$ 
(a) and $\tensor{V}$ (b) yield, through eq.  (\ref{eq:Ppredicted}) with quadratic plasticity function $h$, the theoretical prediction of  $\tensor{P}$ (c), in good agreement with its measurement (d). 
Scale: for $\tensor{U}$, bar$=1$ (no unit); for   $\tensor{V}$ and $\tensor{P}$: 
bar$=1$ s$^{-1}$. 
\label{cap:plasticpredictedBenjamin} }
\end{figure}

For a wet foam, upon flowing around the obstacle
(Fig. \ref{cap:plasticpredictedBenjamin}), we observe that the amplitude of $\tensor{U}$ increases then decreases, while 
$\tensor{V}$ changes orientation. $\tensor{U}$ ellipses have a "capsule" shape before the obstacle, with a negative compression larger than the positive expansion. This is due to bubble compressibility: in this experiment bubbles  can expand their height (in the water bath) and thus retract their horizontal area \cite{dol05PRE}.
 
The agreement between prediction (Fig. \ref{cap:plasticpredictedBenjamin}c) and measurement
 (Fig. \ref{cap:plasticpredictedBenjamin}d) of $\tensor{P}$ is very good.
In particular, we predict well the spatial distribution of T1 events: they occur
mostly just before the obstacle, and in a narrower region after it.
For horizontal positions just on the right of the center of the obstacle,   
the flow tends to decrease the existing strain ($\tensor{V}:\tensor{U}_{d}<0$):
 the predicted frequency vanishes.

We also predict well the direction of rearrangements,
as well as their amplitude, represented by the direction of the coffee
beans and their size, respectively. We do not observe in experiment purely elastic areas and purely plastic areas with a sharp transition line.
This is why  a discontinuous  plasticity function $h$ would be unsuitable, and we use a smooth one (eq. \ref{eq:Ppredicted}).
  
\begin{figure}[hbtp]
 \includegraphics[scale=0.57]{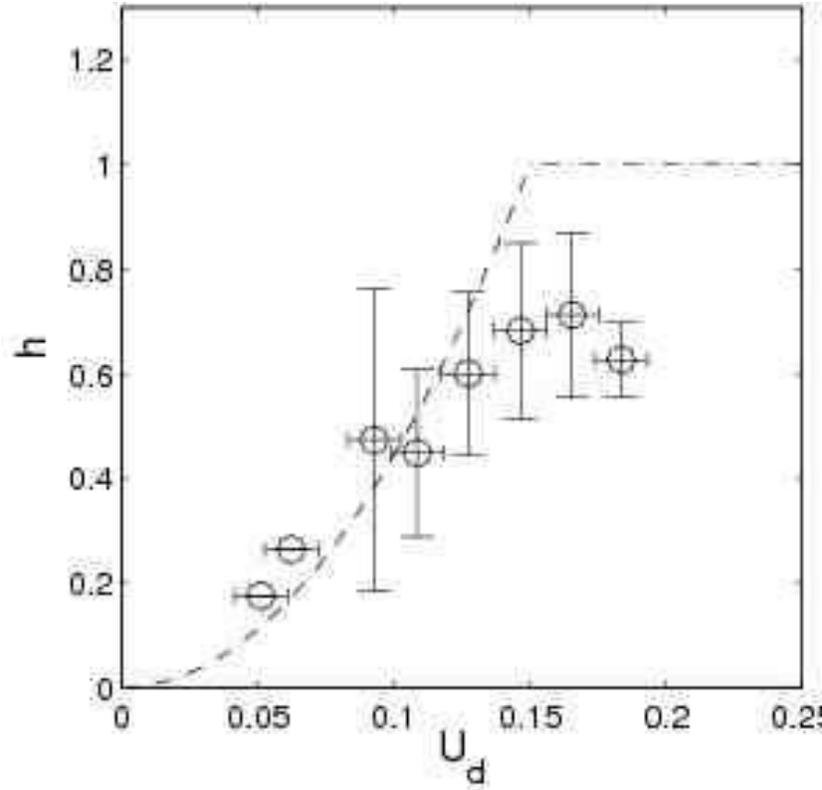} 
  \caption{Estimate of $h$. Symbols: measurements from Fig. (\ref{setups}a) of $ P_{d}/V_{d}\simeq h$. We represented the average (circle) and standard deviation (vertical error bar) after binning data along the horizontal axis on  equal interval sizes (horizontal error bar). Dash-dots:  $h(U_{d})=\min((U_{d}/U_Y)^2,1)$, quadratic up to $U_Y=0.15$.}
  \label{extract_h_benjamin}
\end{figure}

Conversely, statistics are just good enough that we can extract $h$ from measurements.
 (Fig. \ref{extract_h_benjamin}). We observe that $h$ increases, more or less like the proposed   $(U/U_Y)^2$, up to  $U_Y=0.15$, then saturates. Interestingly, it plateaus at a value $\sim 0.6\pm 0.1$ significantly lower than 1. This is probably because after the obstacle $\tensor{V}$ unloads $\tensor{U}$, which  decreases before the foam enters the fully plastic regime.

\clearpage

\subsubsection{Flow around an obstacle: dry foam}

The same experiment with a dry foam (Fig. \ref{setups}b) yields a qualitatively similar behaviour for $\tensor{U}$, $\tensor{V}$ and  $\tensor{P}$ (Fig. \ref{cap:plasticpredictedChristophe}). Quantitatively, however, the maximum value of $U_d$ is here 0.45, which is a reasonable value for a dry foam  \cite{prin83}. The spatial variation of  $\tensor{U}$,  $\tensor{V}$ and  $\tensor{P}$ is restricted to a much narrower range. Note that ellipses are more circular: foam is much less compressible than the previous dry case.

\begin{figure}[hbtp]
 \includegraphics[scale=0.57]{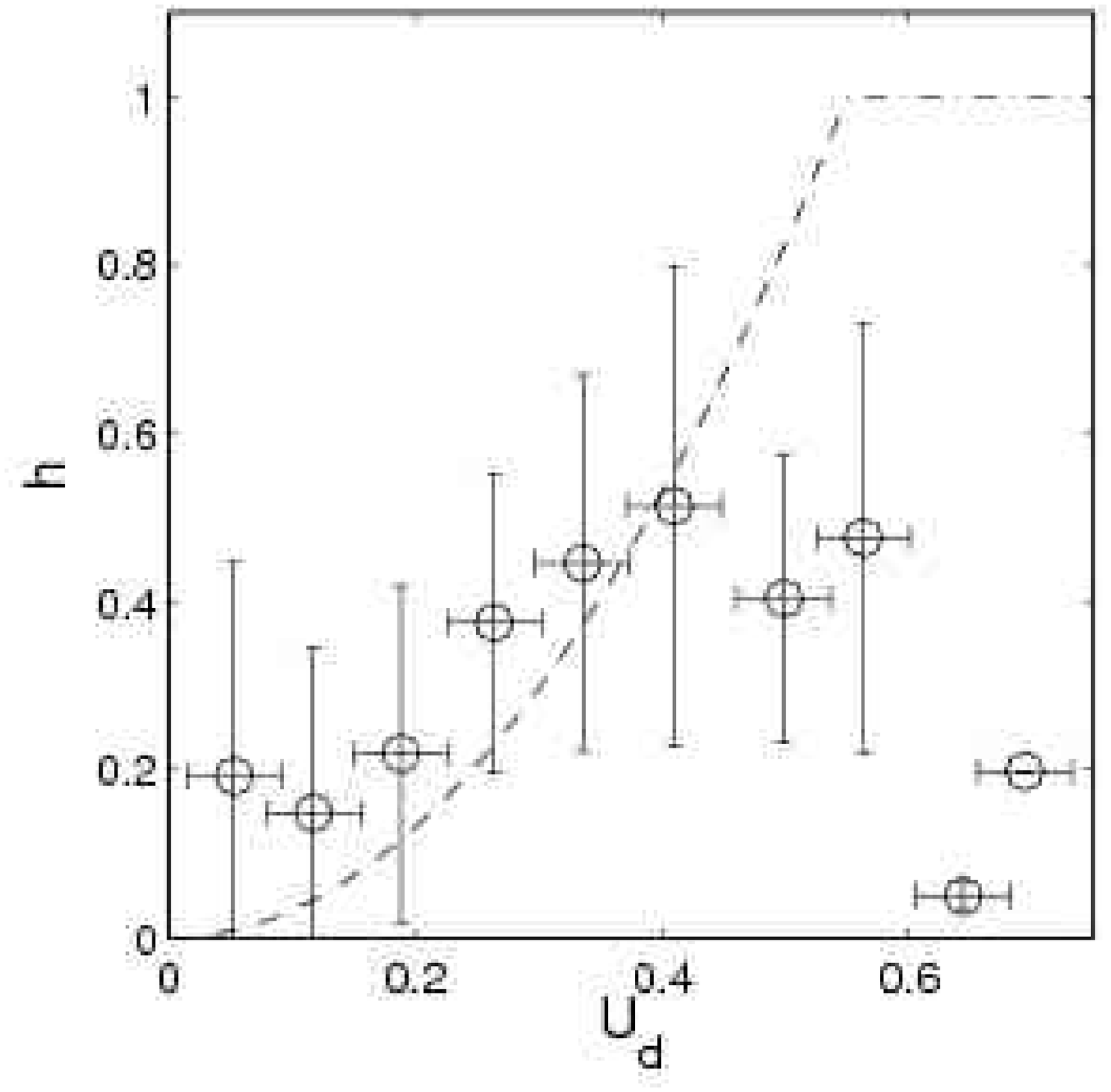}
  \caption{
  Same figure as  Fig. (\ref{extract_h_benjamin}), but for the dry foam of Fig. (\ref{setups}b), $U_Y=0.45$.  }
\label{extract_h_Christophe} 
\end{figure}

\begin{figure}[hb]
 (a)
 \includegraphics[width=6.5cm]{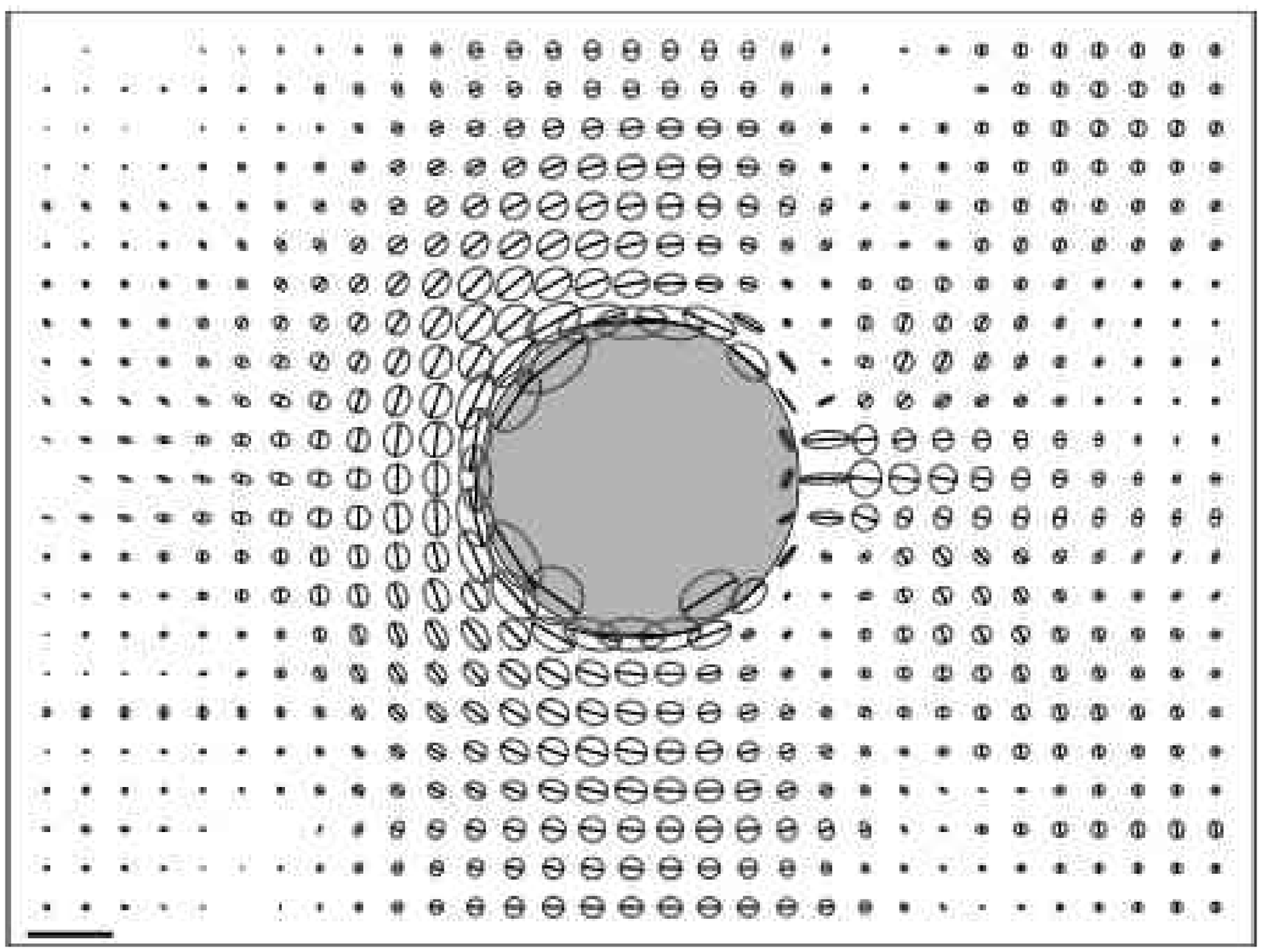}\\
 (b)
 \includegraphics[width=6.5cm]{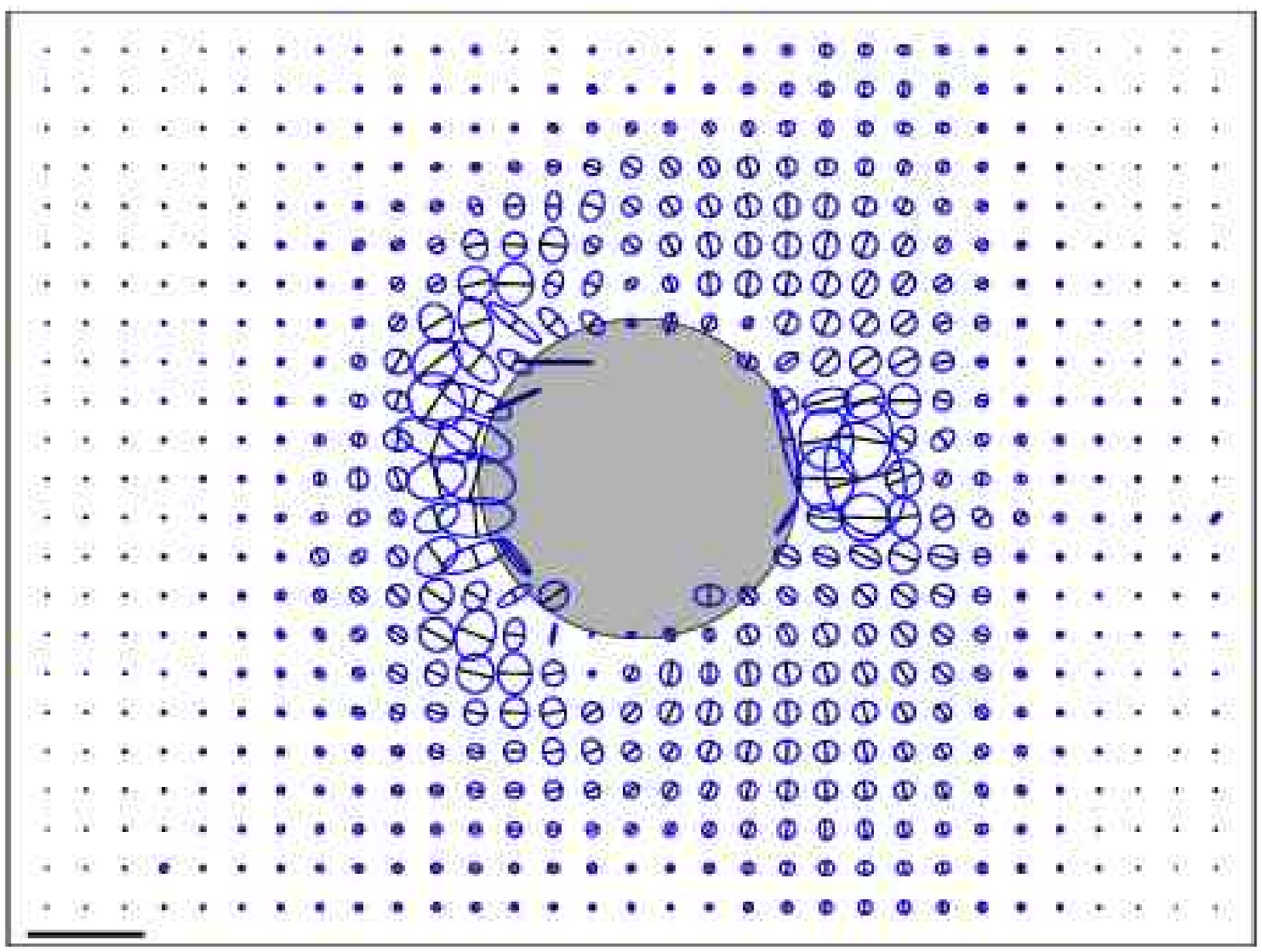}\\
 (c)
 \includegraphics[width=6.5cm]{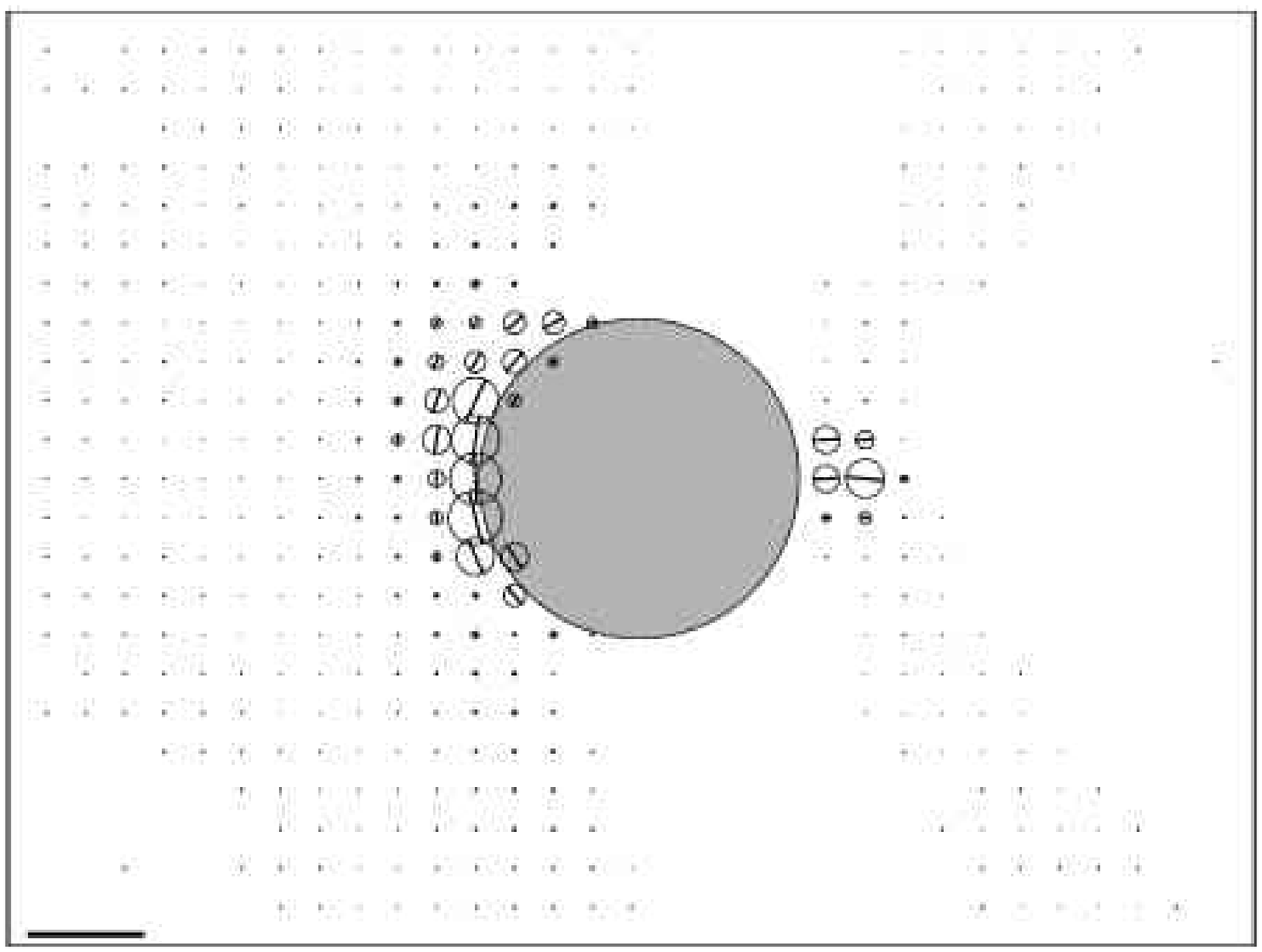}\\
(d)
 \includegraphics[width=6.5cm]{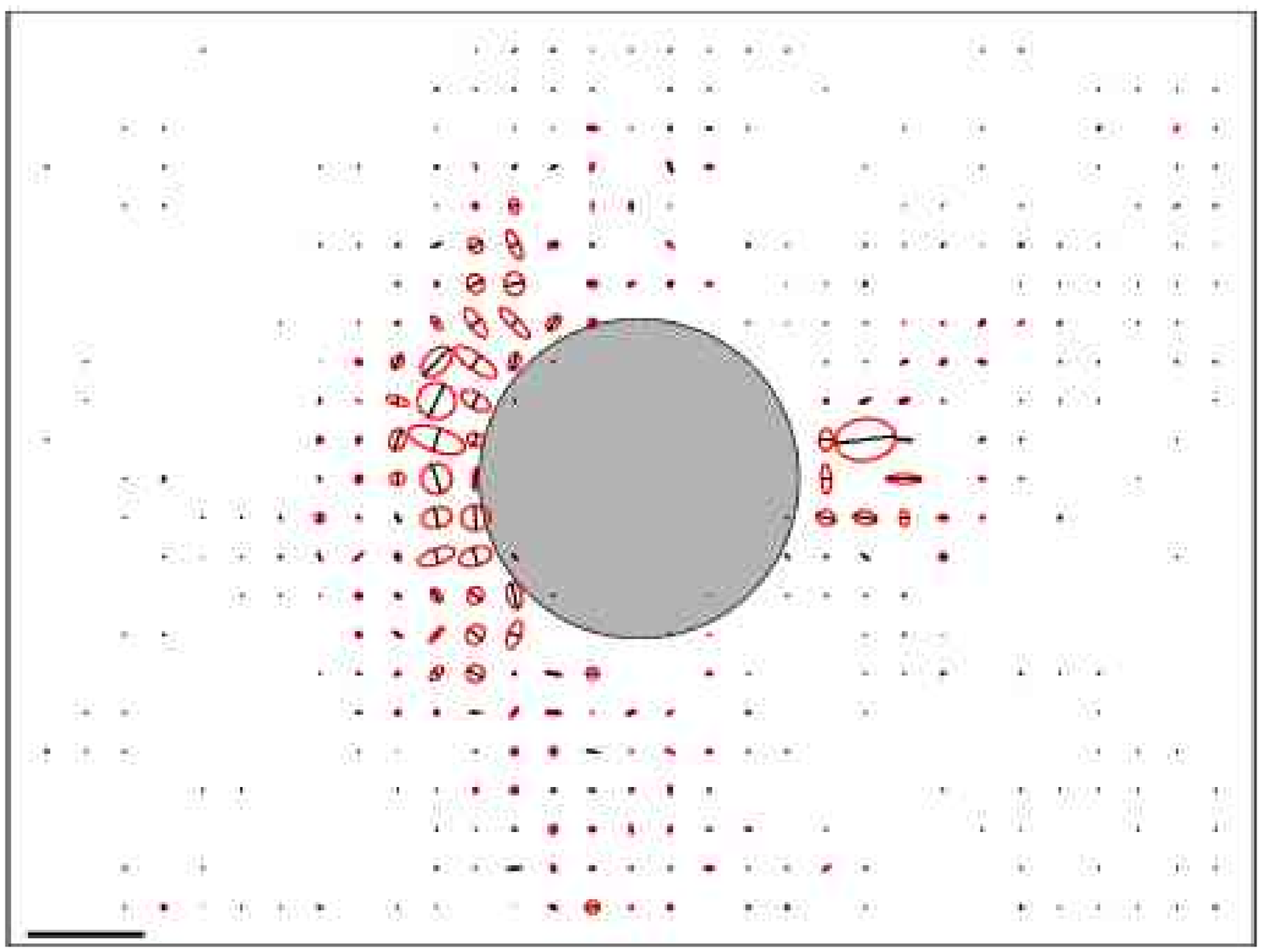}
\caption{Dry foam flowing around an obstacle (Fig. \ref{setups}b). 
Same caption as Fig. (\ref{cap:plasticpredictedBenjamin}), except that
 for   $\tensor{V}$ and $\tensor{P}$  
bar$=0.063$ s$^{-1}$. See similar figures in the companion paper \cite{gra07}.
\label{cap:plasticpredictedChristophe} }
\end{figure}

This means that we   measure larger values but on much less points, resulting in poorer statistics. Still, the agreement between prediction and measurements of $\tensor{P}$ is qualitatively correct. Extracting $h$ from the data is also qualitative, with apparently a plateau as low as 0.4 (Fig. \ref{extract_h_Christophe}). 

\clearpage

\subsubsection{Flow through a constriction. }

\begin{figure}[hb]
 (a)
\includegraphics[height=3.7cm,angle=90]{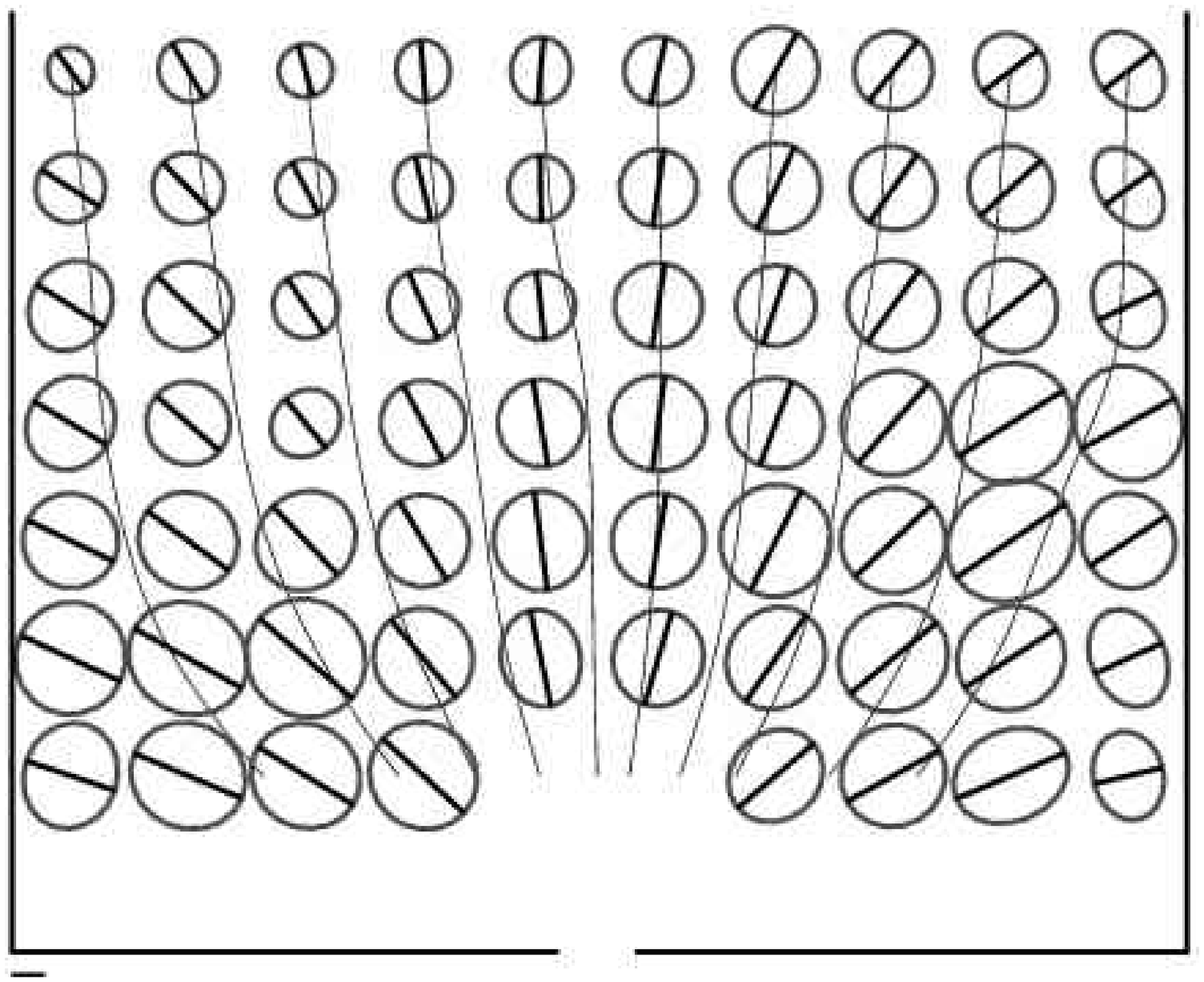}
 (b)
 \includegraphics[height=3.7cm,angle=90]{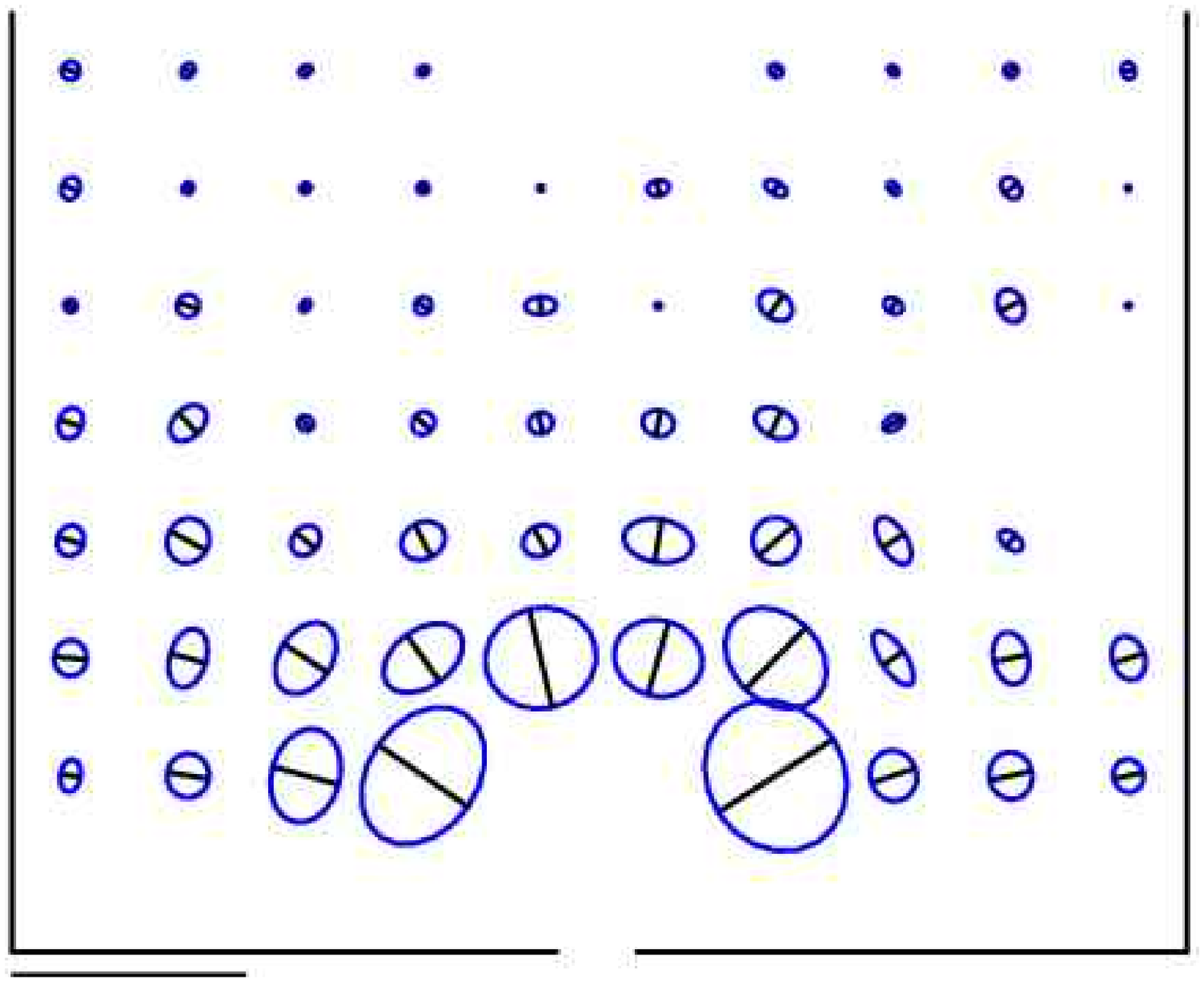}\\
 (c)
\includegraphics[height=3.7cm,angle=90]{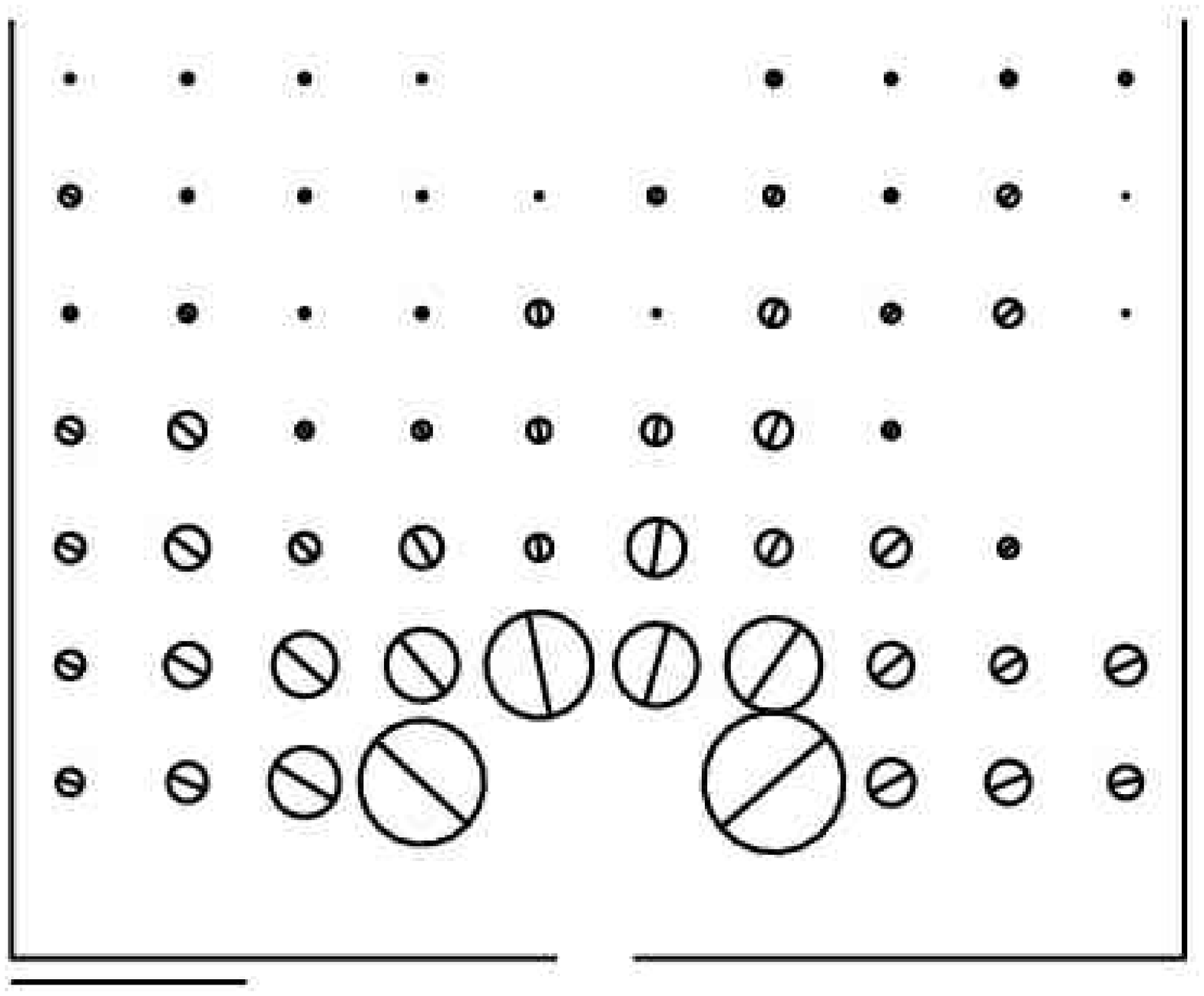}
(d)
\includegraphics[height=3.7cm,angle=90]{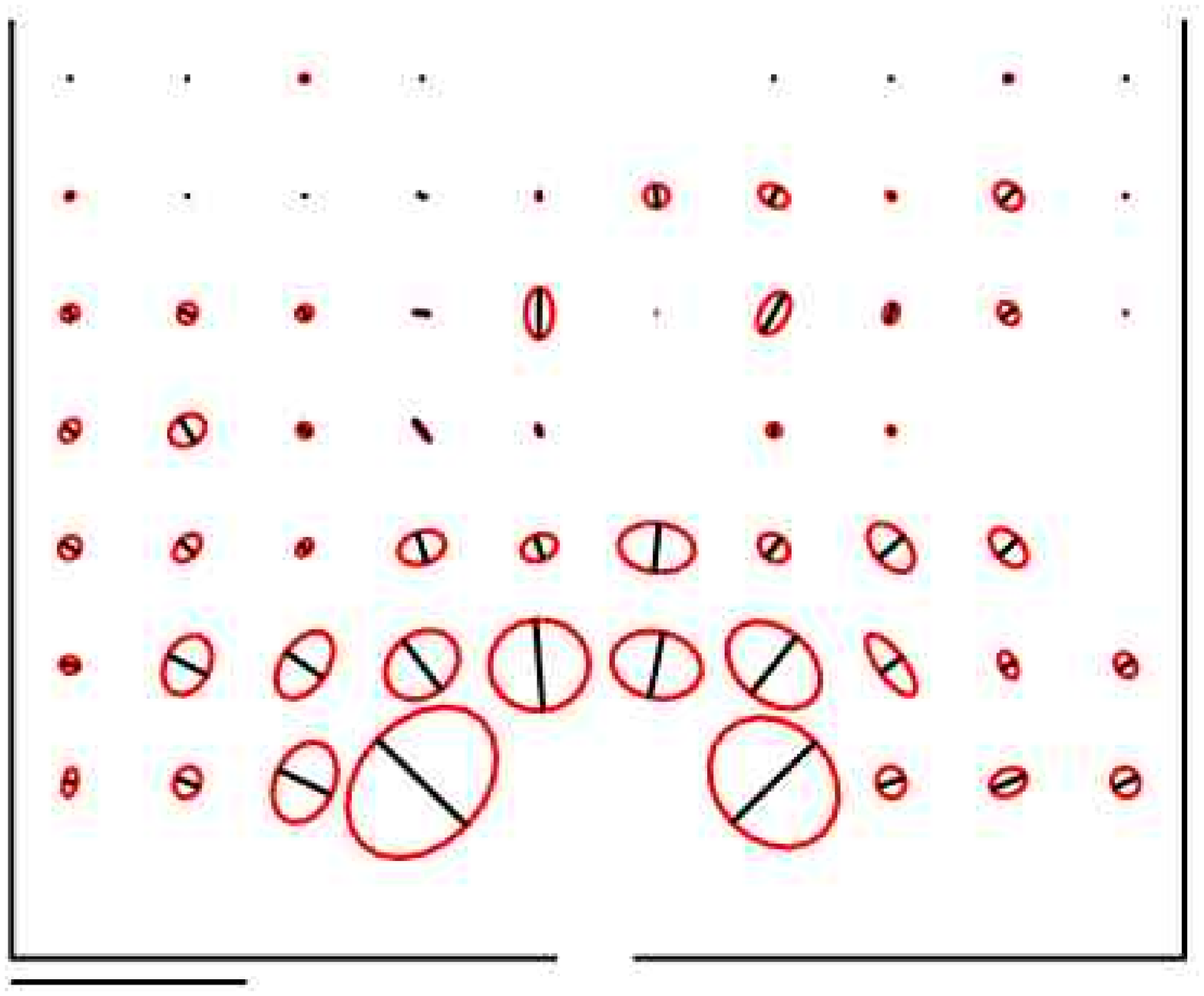}
\caption{Foam flowing through a constriction (Fig. \ref{setups}c). 
Same caption as Fig. (\ref{cap:plasticpredictedBenjamin}), except that
 for   $\tensor{U}$ bar$=0.1$, and  for   $\tensor{V}$ and $\tensor{P}$  
bar$=0.25$ s$^{-1}$. See similar figures in the companion paper \cite{gra07}.
}
\label{fig:CartesMarius}
\end{figure}

When a foam is forced to flow through a  constriction (Fig. \ref{setups}c), along any streamline  $U_{d}$ steadily increases. The constriction is here so narrow, comparable to the square root of a bubble area, that just at the constriction the continuous description (and thus our measurement method) breaks down.
The influence of the constriction is visible far uphill:   $U_{d}$ is widely distributed, and we obtain good statistics (Fig.  \ref{fig:CartesMarius}).

\begin{figure}[hbtp]
(a)
 \includegraphics[scale=0.57]{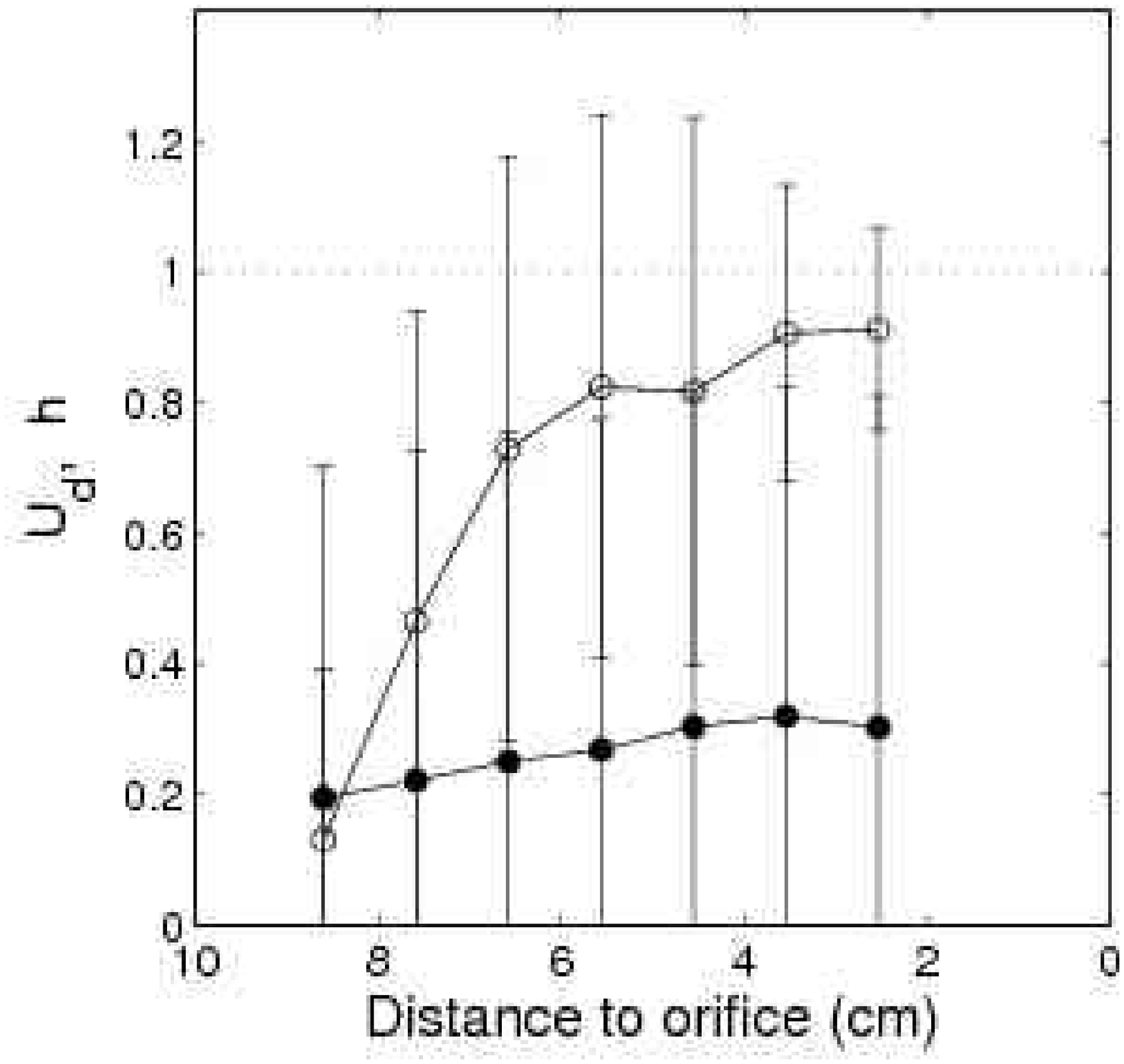}\\
  (b)
  \includegraphics[scale=0.57]{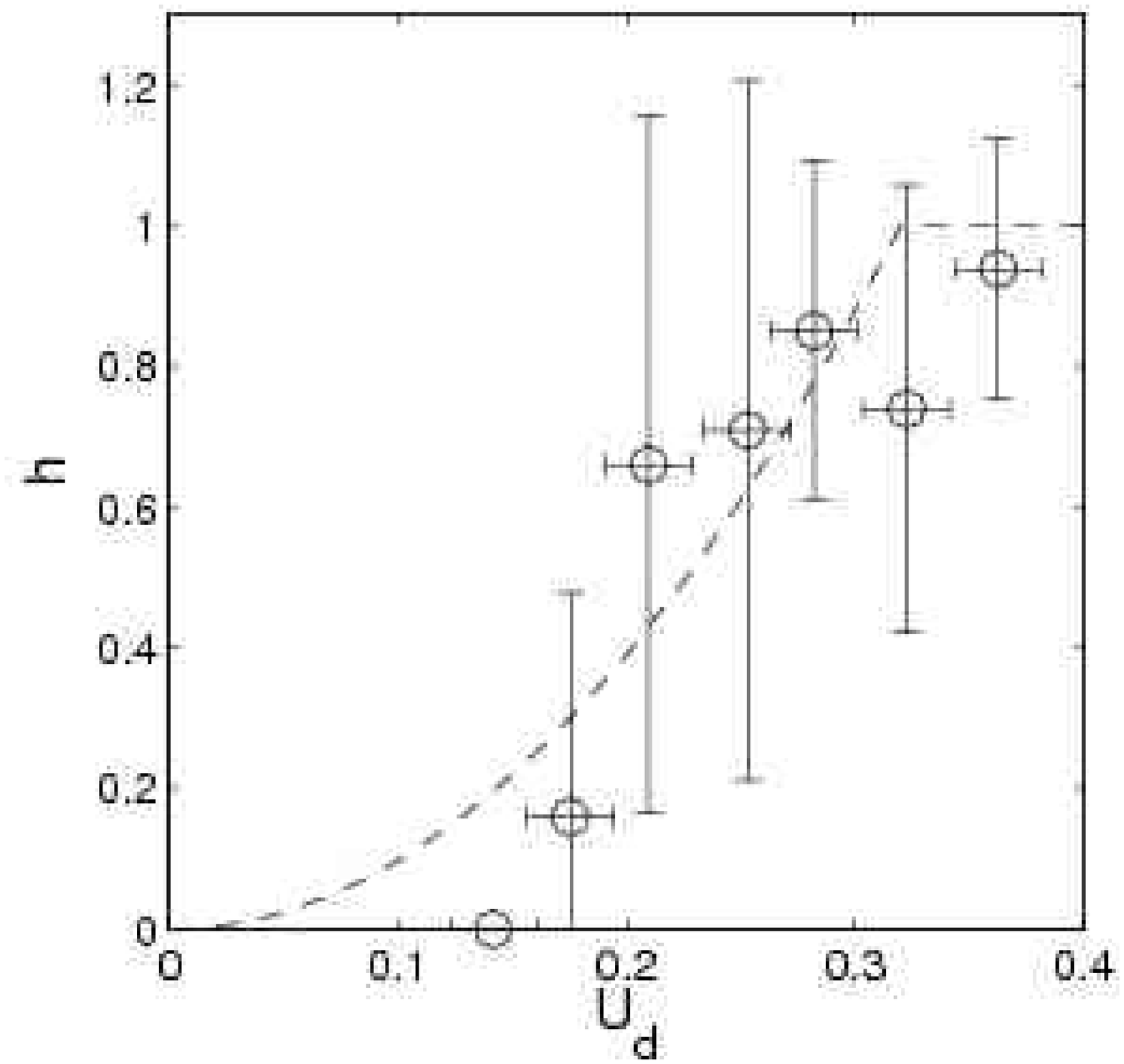}
  \caption{
  Constriction: analysis of Fig. (\ref{setups}c). 
  (a) $  P_{d}/V_{d}$ (open circles) and $U_{d}$ (closed circles) {\it versus}  the distance to the constriction. (b) Plasticity fraction $h$ estimated as $h\simeq P_{d}/V_{d}$  {\it versus } strain
  $U_{d}$ (data from (a)): same figure as  Figs. (\ref{extract_h_benjamin}) and (\ref{extract_h_Christophe}), here with 
  $U_Y=0.32$.
    }
  \label{fig:hMarius}
\end{figure}

Conversely,  $\tensor{V}$ is more localized near the orifice, and thus, as expected from 
eq.  (\ref{eq:Ppredicted}), so is   $\tensor{P}$.
Plasticity is indeed oriented by the elastic strain, the angle between the main axis of $\tensor{P}$ and
 of $\tensor{U}_{d}$ is $1\pm5^{\circ}$.  
Concerning the plasticity amplitude, we  observe that $P_d$ is much smaller than $V_d$ when the foams enters in the field of view:
$P_d/V_d$ undergoes a 5-fold increase until it reaches 1 at the constriction  (Fig. \ref{fig:hMarius} a), and 
$U_d$ plateaus. We measure $U_Y\sim 0.32$; which is reasonable for a foam with $\sim$1\% liquid fraction.

The direct estimate of the plasticity fraction as  $h\simeq P_{d}/V_{d}$  is obtained  with reasonably good statistics (Fig. \ref{fig:hMarius} b).  It plateaus close to 1, confirming that the saturation is reached. 
We obtain an extremely good quantitative agreement of the predicted plasticiy with the measurement (Fig.  \ref{fig:CartesMarius}).

\subsubsection{Couette flow }

 \begin{figure}[t]
 (a)
 \includegraphics[height=3.7cm,angle=180]{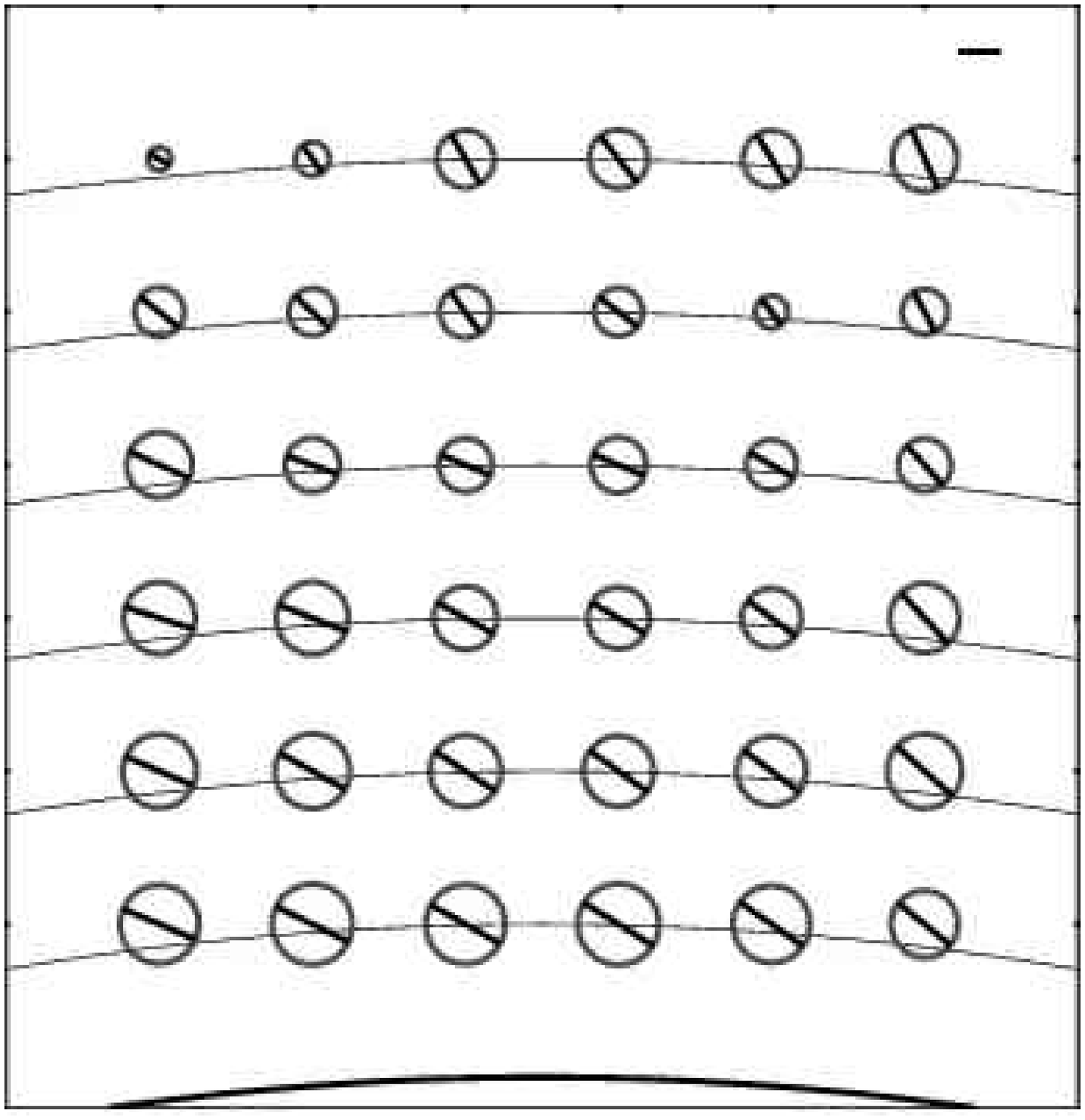}
 (b)
 \includegraphics[height=3.7cm,angle=180]{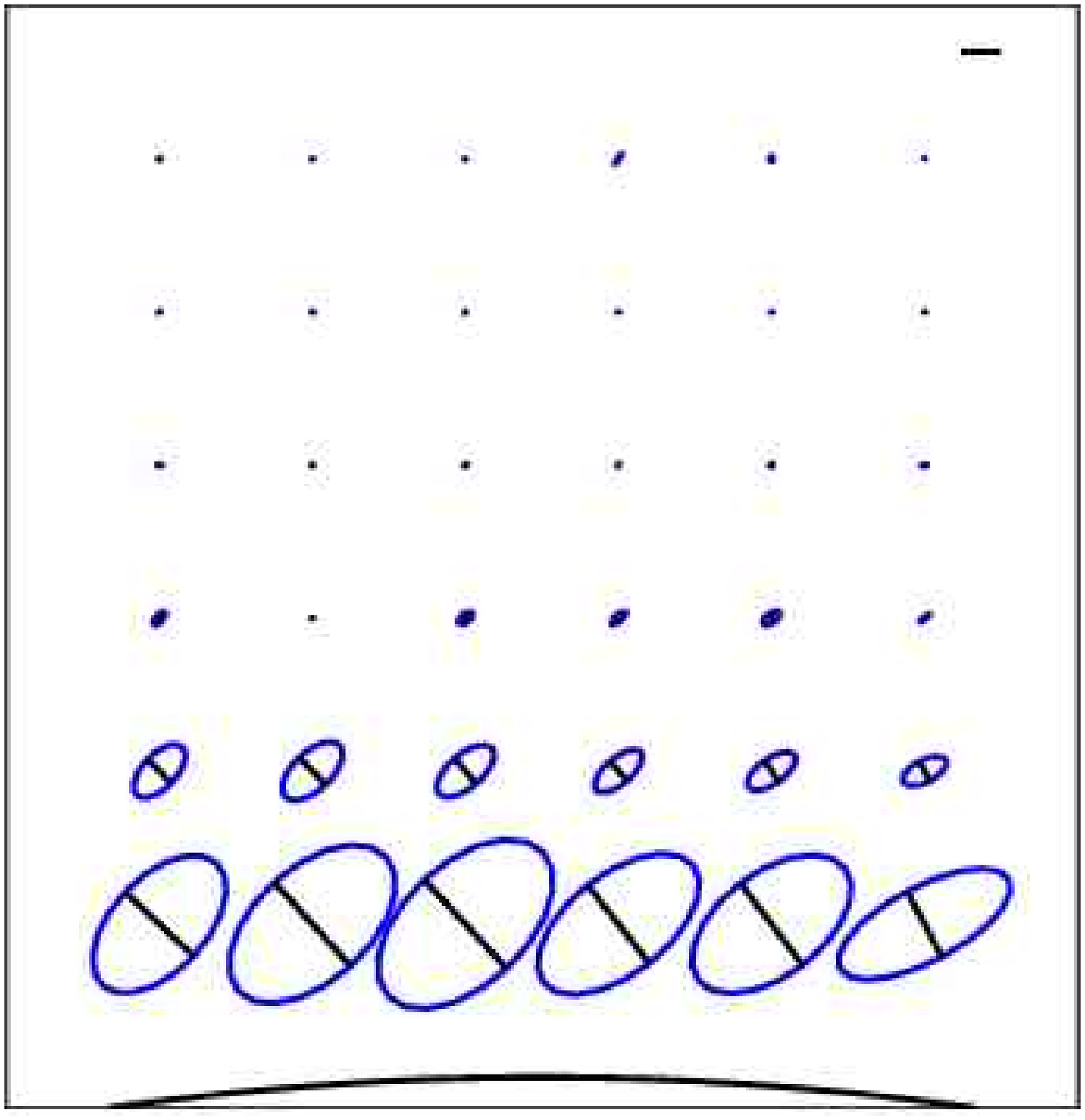}
 \\
 (c)
\includegraphics[height=3.7cm,angle=180]{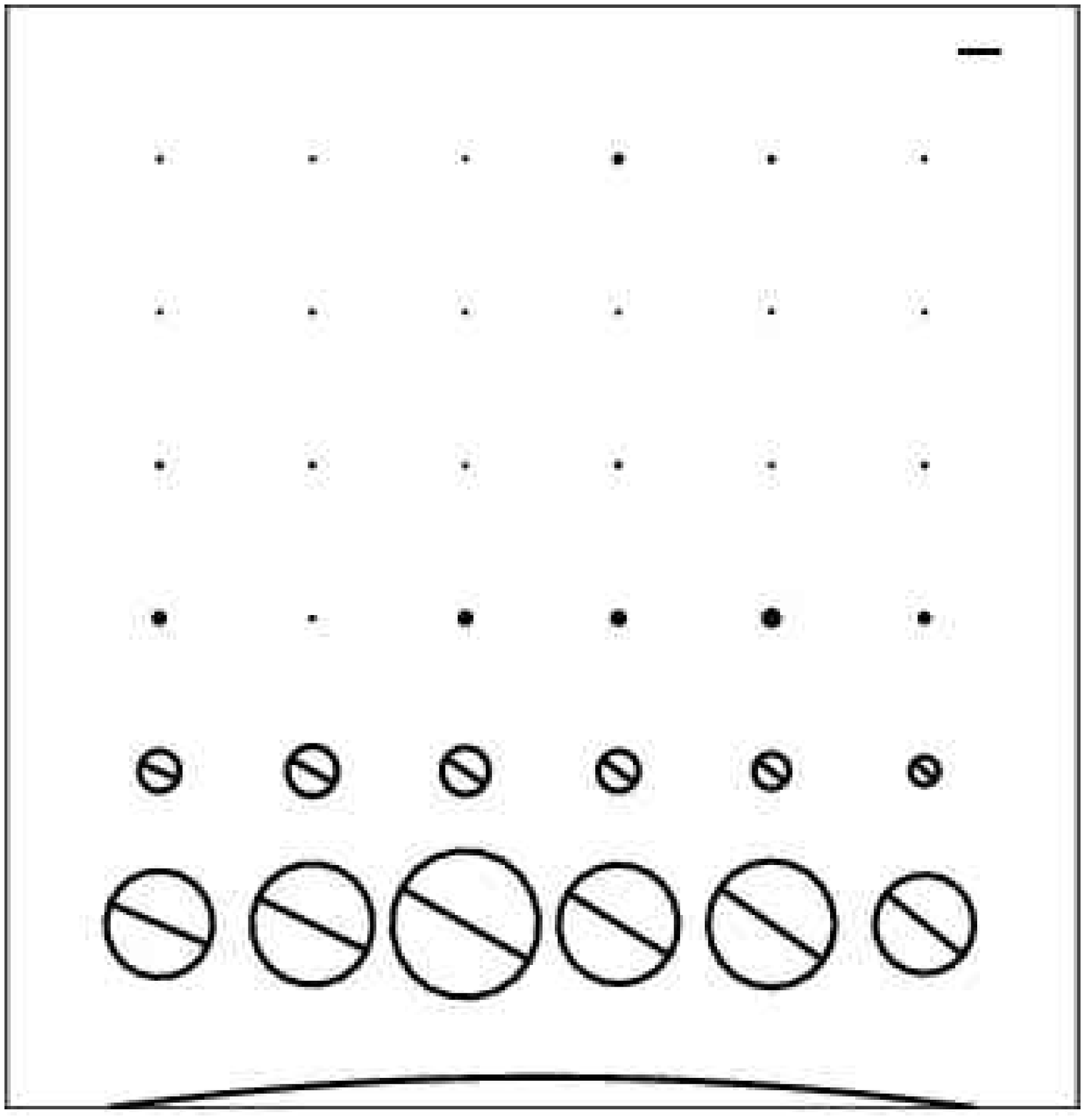}
(d)
\includegraphics[height=3.7cm,angle=180]{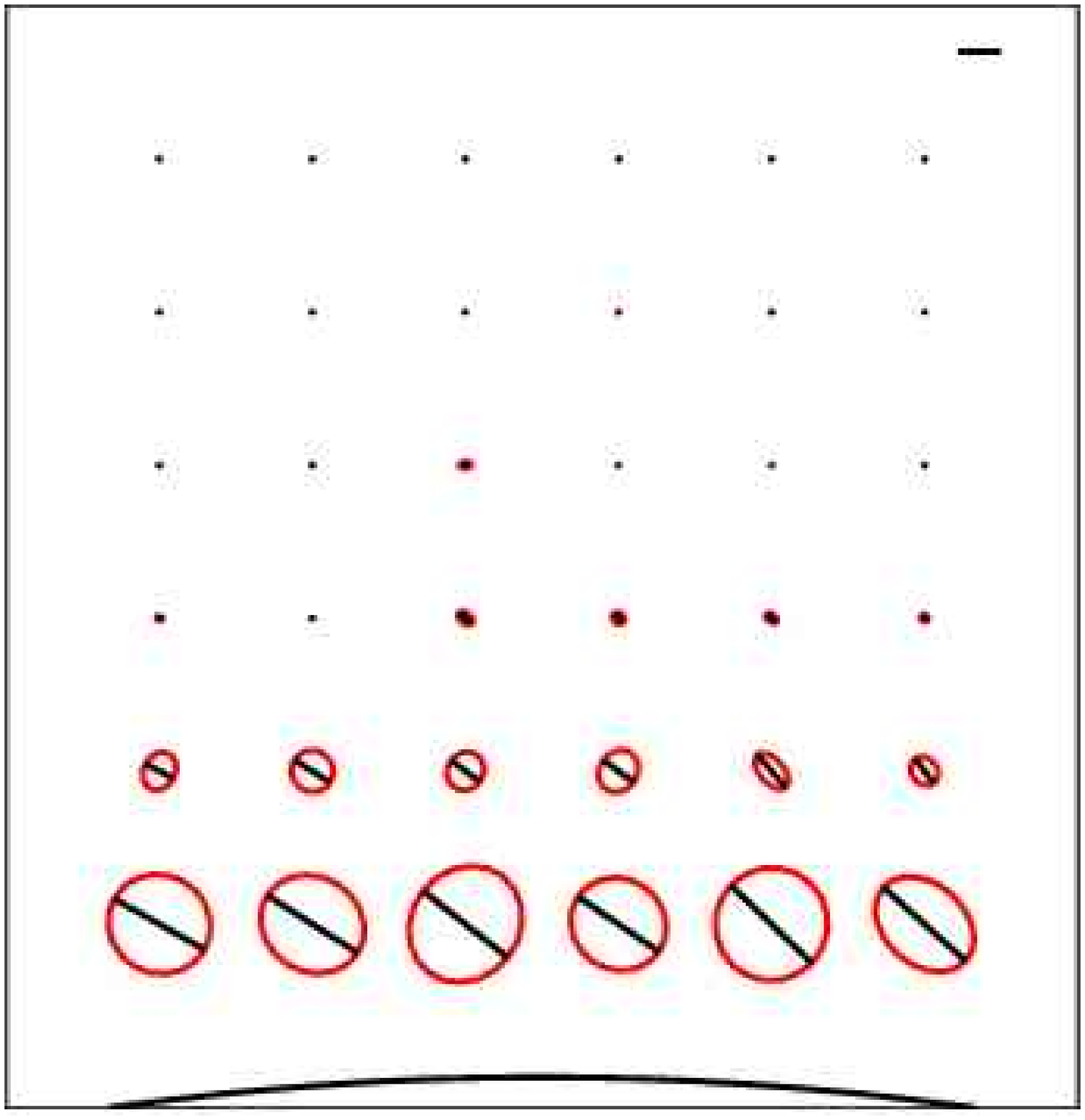}
\caption{
Foam sheared between concentric disks (Fig. \ref{setups}d). 
Same caption as Fig. (\ref{cap:plasticpredictedBenjamin}), except that
 for   $\tensor{U}$ bar$=0.1$, and  for   $\tensor{V}$ and $\tensor{P}$  
bar$=0.25$ s$^{-1}$. 
}
\label{fig:CartesGeorges}
\end{figure}

In a steady Couette flow (Fig. \ref{fig:CartesGeorges}), as expected $\tensor{U}$ respects the circular symmetry: it does not vary orthoradially. The advantage is that we can improve the measurements by averaging orthoradially. The drawback is that we have very few independent measurements (here 6), along the radial direction. $\tensor{U}$ is significantly different from zero everywhere. Near the rotating (inner) disk, it means that $\tensor{U}$  saturates. Near the 
  fixed (outer) disk, it is probably  a residual strain due to the foam preparation (there are not enough T1s to relax it). $U_d$ is rather low (at most of order of 0.1), which is consistent with the high liquid fraction. All these findings confirm those of ref. \cite{jan05}.
  
As expected, $\tensor{V}$ similarly does not vary orthoradially. It decreases quickly with the distance to the fixed disk, so that we have only two independent, non-zero measurements. It is thus impossible to perform the same analysis as in the above   flows which truly vary with both space coordinates. 
   
Still, we can predict    $\tensor{P}$ from eq.  (\ref{eq:Ppredicted}): this agrees quantitatively with the measurements. 
Concerning the orientation of $\tensor{P}$, the angle between the main axis of $\tensor{P}$ and of
 $\tensor{U}$ is $-5\pm3^{\circ}$, they are indeed aligned.
Since the flow is steady, $\tensor{U}$  is constant along a streamline, and we thus expect 
    $\tensor{P}=\tensor{V}$: this agrees qualitatively with  the measurements.


\section{Conclusion}
\label{concl}

\subsection{Summary}
 
Using the formalism developed in the companion paper \cite{gra07}, we measure in different 2D foam flows the matrices which quantify   the elastic strain, the total strain rate and  the  plastic rearrangements. We identify statistical measurements performed on the detailed structure of bubbles, with large scale quantities describing the foam as a continuous material.

We then generalize to matrices a previous scalar model \cite{pinceau}, and base it on local measurements on individual bubbles. 
We show that the  plastic rearrangements arise from a combination of
both the elastic strain and the total strain rate. As shown by the maps, they cannot be predicted from the  elastic strain alone, nor from the total strain rate alone. 

For instance, in the wet obstacle flow, the spatial symmetry with respect to the obstacle is very different in the three maps. In the dry foam obstacle, the elastic strain extends very far, while the velocity gradient has a narrower extension, but both are needed, and the total strain gives the orientation. In the constriction the elastic strain extends so much that it is mainly the total strain rate which determines where T1s occur.

In a first approximation, the plasticity is described mainly by the behaviour near yielding. The yield strain $U_Y$  is the main relevant parameter. We determine it directly from image analysis and check that the obtained values are reasonable. 
We then statistically predict the position, orientation, anisotropy and frequency of topological rearrangements in a flowing foams, in good agreement with various experiments. 

In a second step, to refine the  description and improve the prediction, we introduce the proportion of plasticity, to account for the gradual appearance of plasticity instead of a sharp yield. It is a function $h(U_d)$ (also called "yield function") of the elastic strain, which interpolates between 0 (fully elastic)  at small strain $h(0)\sim 0$, and 1  (fully plastic) near yielding, $h(U_Y)\sim 1$. We obtain here estimates of $h$.
 
\subsection{Discussion}
\label{discussion}
 
The model  presented in Section \ref{plasticmodel} succesfully predicts the plasticity,
and contributes to describe the foam as a continuous material.
We now discuss its validity and some of its limitations.

\subsubsection{Material}

The general approach which links individual and collective descriptions is applicable to other complex materials (see companion paper \cite{gra07}). The identification between statistical and continuous quantities is restricted to affine flows, which probably applies rather generally to foams and  other cellular materials; it also requires that the elastic strain corresponds directly to individual objects, 
and the plastic strain rate to topological rearrangements: this is probably specific to dry foams and emulsions.

\subsubsection{Shear rate}

Foams coarsening, due to the diffusion of gas from one bubble to the other, couples to foam rheology at slow time scales \cite{hoh05}. Here, the present experiments are short enough to neglect this effect. 
  
On the other hand, at very high shear rate, the rheology couples to the bubble's internal relaxation times, which are much shorter   \cite{Saramito2007}. In a material like a foam, at such a high shear rate the bubbles' films and vertices do not relax towards the local mechanical equilibrium. Bubbles distort and the foam eventually breaks; in fact, at all velocities explored here, if the foam exists it means that the shear rate is slower than the internal relaxation times \cite{rau07}.

\subsubsection{Dimension}

Our  formalism is written using matrices, so that it 
indifferently applies to 2D or 3D systems. 
In principle it could be tested in truly 2D foams (for instance in numerical simulations) as well as in truly 3D foams (detailed measurements are in progress). 
Here, for simplicity, 
we test it  on bubble monolayers (quasi-2D foams) which flow horizontally (true 2D velocity field).

The friction due to plates of glass is an external force acting on bubbles, not to be confused with viscous stresses which are internal to the foam. It likely plays a role in the values of the quantities we measure above, but does not affect the {\it relations} between these quantities; that is, the equations we write are insensitive to this friction. 
This is in particular the case for  the repartition of the total strain rate between its   elastic and plastic contributions (eq. \ref{eq:Ppredicted}). 
\label{frictionplates}  

\subsubsection{Disorder}

We have assumed here that the foam is amorphous; that is, RVEs are homogeneous. 
This excludes  foams with: crystalline order;  avalanches of T1s; localization of the shear, and shear banding; or  one dimension comparable to a bubble diameter.

 Here we measure the effects of the function $h$, and thus indirectly trace back to its approximate expression. This is not precise, since apparently, the exact shape of $h$ does not affect much the predictions. This implies that, reciprocally, once $h$ is known, predictions based on it are very robust. 
We are thus currently trying to measure  directly the function $h$ for various foams,  with different liquid fractions and disorders.

\subsection{Perspectives: constitutive equation}
\label{constit}

\subsubsection{Closing the set of equations}

In principle we could study the fluctuations in time and space of the three matrices we measure: $\tensor{U}$, $\tensor{P}$, $\tensor{V}$. However, we observe experimentally that their averages  vary smoothly with space. 
As a first step, we want to use these average smooth fields  to obtain a continuous description  of  2D foam flows. For that purpose, we need a closed set of equations  to relate these three matrices.

First, the kinematic equation  (\ref{eq:dUdt}) means that the total strain rate  is shared between elastic and plastic contributions.  It is equivalent to rewrite it as:
\begin{equation}
\frac{{\cal D}{\tensor{U}}}{{\cal D}t}=
\tensor{V}
-
\tensor{P}.
\label{eq:VpourdUdt}
\end{equation}
In other words,  the elastic strain is loaded by the total applied strain rate, that is, the velocity gradient; this  process   is   limited by   the plastic strain rate, which saturates the elastic strain.

Second, the plasticity  equation   (\ref{eq:Ppredicted})  describes how the total strain
rate is shared between change of elastic strain and plastic strain
rate. It involves the fraction of plasticity, $h$. Eq. (\ref{eq:Ppredicted}) can be combined with eq. (\ref{P_partI}) into a single equation,
using the Heaviside function:
\begin{equation}
\tensor P=  
{h(U_{d})}\; {\cal H}(\tensor{V}\! : \! \tensorstrain_{d})  \; 
 \frac{({\tensor{V}\! : \! \tensorstrain_{d}})
 \:\tensorstrain_{d}}{2{U_{d}}^2} .
\label{eq:PpredictedHeaviside} 
\end{equation}

Third,  $\tensor{V}$ is directly related to the symmetrised gradient of the velocity field. In fact, combining eqs. (\ref{G_affine}) and (\ref{definiV}) yields:
 \begin{equation}
\tensor{V}  
\;    \stackrel{\rm affine}{\simeq}\; 
 \frac{ \tensor{\nabla v}+ \tensor{\nabla v}^t}{2}.
 \label{V_affine}
\end{equation}

Fourth, the velocity field is determined by the fundamental equation of dynamics, that is, momentum conservation (Navier-Stokes):
 \begin{equation}
 \rho \left( \frac{\partial \vec{v}}{\partial t} + (\vec{v}.\nabla)\vec{v}\right) = 
 {\rm div}.\tensor{\sigma} + \vec{f}_{ext}
  \label{Navier-Stokes}
\end{equation}
where $\rho$ is the foam density, $\tensor{\sigma}$ the stress tensor, and $ \vec{f}_{ext}$ the external forces (such as the friction on the glass plates) not included in the stress.

Fifth,  we  now need the constitutive equation itself. That is,   a dynamical equation  which determines the elastic
and viscous contributions to the stress, from the current elastic
strain and strain rate.
We suggest to generalize eq. (\ref{sigma_pinceau}), now using matrices:
\begin{equation}
\tensor{\sigma}
=2 G \tensor{U}_{d}
+ K \: {\rm Tr}(\tensor{U})\tensor{I}
+ 2\eta \tensor{V}.
\label{eq:DynamicalEquation}
\end{equation}
We have added here  a compression modulus $K$;  in practice, for foams, it is  much larger than the shear modulus. 
We write the material-dependent parameters as scalar, which is correct for an isotropic disordered foam; in an anisotropic material they would be rank-four tensors.
The drawback of the external friction on plates (mentioned in section \ref{frictionplates}) is that it prevents us from measuring locally the viscous contribution to the stress in eq.  (\ref{eq:DynamicalEquation}). We thus have assumed its viscosity is linear, {\it i.e.} $\eta$ constant. It is possible in principle that $\eta$ depends on $\tensor{U}_{d}$ or $\tensor{V}$, thus
 introducing some non-linearities. 
 
In conclusion,  this fully closed set of equations can be solved
 from the knowledge of two  field variables: $\vec{v}$ for transport and $\tensor{U}$ for internal strain.
   

\subsubsection{Unified description of elastic, plastic and viscous behaviours}

Together, eqs. (\ref{eq:VpourdUdt}-%
\ref{eq:DynamicalEquation})  provide a fully closed system. 
 It is a complete continuous constitutive equation for foams.
It treats with an equal importance the elastic, plastic and viscous contributions in any of the three regimes:
elasticity, plasticity and flow. It admits as limits the classical equations of elasticity or hydrodynamics, as well as elasto-plasticity, visco-elasticity and visco-plasticity. Extension to higher shear rates  \cite{Saramito2007} is challenging,  but not impossible in principle. 

It should be enough to perform complete, testable predictions of  $\tensor{U}$, $\tensor{P}$, $\tensor{V}$, from the material's properties and the boundary conditions. This is much stronger than 
Sections \ref{plasticmodel} and \ref{tests}, which predict  $\tensor{P}$ from known $\tensor{U}$ and $\tensor{V}$. Such analytical predictions, and the corresponding experimental tests, are in progress.  They are outside of the scope of the present paper; still, we can make a few remarks regarding the foam dynamical properties.
  
\begin{figure} 
\begin{center}
\includegraphics[width=7.8cm]{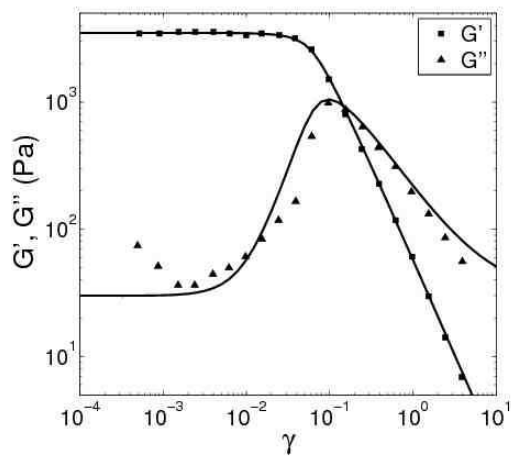}
\caption{Storage and loss moduli versus strain amplitude for a
monodisperse emulsion. Symbols: experimental $G'$ (squares) and
$G''$ (triangles) in a close-packed emulsion (Fig. 1 of Ref. \cite{Mason1995}). Lines: model calculated with the complete set of equations, 
eqs. (\ref{eq:VpourdUdt}-%
\ref{eq:DynamicalEquation}).
The method is the same as in  Ref. \cite{pinceau};
the correspondence between scalars and matrices is obtained by replacing ($\sigma$,$U$,$\dot{\varepsilon}$) by ($\sigma_{xy}$,$2U_{xy}$,$2V_{xy}$).
The factors 2 come from the tensorial generalization (eq. (\ref{eq:DynamicalEquation})) of the scalar equation (eq. (\ref{sigma_pinceau}));
this is why the yield strain is half of that used for the scalar case (Fig. 9 of Ref. \cite{pinceau}).
The parameters are directly obtained from the experimental data:
shear modulus $G = 1.7 103$ Pa, yield strain $U_Y = 0.0223$, 
and
viscosity $\eta = 30$ Pa.s. 
The
yield
function used here is $h = (U/U_Y)^2$.
}
\label{Gprimematriciel}
\end{center}
\end{figure}

  In oscillatory regime, it predicts  
the storage and loss moduli $G'$, $G''$  (Fig. \ref{Gprimematriciel}). Since these are scalar quantities, this is very similar to predictions based on eq. (\ref{sigma_pinceau}) \cite{pinceau}.
 
  In stationary regime, while we can not yet measure locally the viscous contributions to the stress, we can indirectly measure its consequences at the global level. In fact, the force acting on a circular obstacle  (Fig. \ref{setups}a) results from the integral of the stress over the obstacle boundary:
  experimentally, it increases linearly with the foam velocity, with a non-zero intercept 
  \cite{dol05PRE}. This agrees with eq. (\ref{eq:DynamicalEquation}).
A consequence is that, in a stationnary regime where the elastic contribution to stress is constant,   
the foam seems to behave as a visco-plastic  (Bingham) fluid.

In any transient regime, however, the full visco-elasto-plastic nature of the foam has to be taken into account. The internal variable $\tensor{U}$, often overlooked,  thus plays a role as important as 
that of $\tensor{P}$ or $\tensor{V}$.

\section*{Acknowledgments}
We warmly thank
B. Dollet, M. Asipauskas, and G. Debr\'egeas for kindly providing published and unpublished raw data.


\begin{thebibliography}{30}

\bibitem{gra07}
F.~Graner, B.~Dollet, C.~Raufaste, P.~Marmottant, companion paper,
  \texttt{http://hal.archives-ouvertes.fr/hal-00160733/en/}

\bibitem{wea99}
D.~Weaire, S.~Hutzler, \emph{The physics of foams} (Oxford University Press,
  Oxford, 1999)

\bibitem{sai99}
A.~Saint-Jalmes, D.~Durian, J. Rheol. \textbf{43}, 6 (1999)

\bibitem{hoh05}
R.~H\"ohler, S.~Cohen-Addad, J. Phys. Cond. Matt. \textbf{17}, R1041 (2005)

\bibitem{pha02}
N.~Phan-Thien, \emph{Understanding viscoelasticity} (Springer-Verlag, Berlin,
  2002)

\bibitem{mac94}
C.W. Macosko, \emph{Rheology : principles, measurements and applications}
  (Wiley-VCH, 1994)

\bibitem{Schwedoff1900}
T.~Schwedoff, \emph{La rigidit{\'e} des fluides}, in \emph{Rapports du
  Congr{\`e}s Intern. de Physique} (1900), Vol.~1, p. 478

\bibitem{White1981}
J.~White, Rheologica Acta \textbf{20}, 381 (1981)

\bibitem{jan06}
E.~Janiaud, D.~Weaire, S.~Hutzler, Phys. Rev. Lett. \textbf{97}, 038302 (2006)

\bibitem{pinceau}
P.~Marmottant, F.~Graner, Eur. Phys. J. E \textbf{23}, 337 (2007)

\bibitem{Saramito2007}
P.~Saramito, Journal of Non-Newtonian Fluid Mechanics \textbf{145}, 1 (2007)

\bibitem{Benito2007preprint}
S.~Benito, C.H. Bruneau, T.~Colin, C.~Gay, F.~Molino, preprint arXiv:0711.0388

\bibitem{deSouzaMendespreprint}
P.R. de~Souza~Mendes, submitted to J. Fluid Mech.

\bibitem{pic04}
G.~Picard, A.~Ajdari, F.~Lequeux, L.~Bocquet, Eur. Phys. J. E
  \textbf{15}(371--381) (2004)

\bibitem{dol05PRE}
B.~Dollet, F.~Elias, C.~Quilliet, C.~Raufaste, M.~Aubouy, F.~Graner, Phys. Rev.
  E \textbf{71}, 031403 (2005)

\bibitem{asi03}
M.~Asipauskas, M.~Aubouy, J.A. Glazier, F.~Graner, Y.~Jiang, Granular Matt.
  \textbf{5}, 71 (2003)

\bibitem{deb01}
G.~Debr\'egeas, H.~Tabuteau, J.M. di~Meglio, Phys. Rev. Lett. \textbf{87}(17),
  178305 (2001)

\bibitem{Raufaste2007}
C.~Raufaste, B.~Dollet, S.~Cox, Y.~Jiang, F.~Graner, Eur. Phys. J. E
  \textbf{23}, 217 (2007)

\bibitem{rau07}
C.~Raufaste, {Ph}D thesis, Univ. Grenoble I (2007),
  \texttt{http://tel.archives-ouvertes.fr/tel-00193248/en/}

\bibitem{jan05}
E.~Janiaud, F.~Graner, J. Fluid Mech. \textbf{532}, 243 (2005)

\bibitem{prin83}
H.M. Princen, J. Colloid Interface Sci. \textbf{91}(1), 160 (1983)

\bibitem{Bragg1947}
W.L. Bragg, J.F. Nye, Proc. R. Soc. London Ser. A \textbf{120}, 474 (1947)

\bibitem{Gouldstone2001}
A.~Gouldstone, K.J. Van~Vliet, S.~Suresh, Nature \textbf{411}(6838), 656
  (2001), ISSN 0028-0836

\bibitem{BubbleRafts}
{DoITPoMS}, {D}issemination of Information Technology for the Promotion of
  Materials Science, University of Cambridge, UK.,
  \texttt{http://www.doitpoms.ac.uk/tlplib/dislocations}

\bibitem{eli99}
F.~Elias, C.~Flament, J.A. Glazier, F.~Graner, Y.~Jiang, Phil. Mag. B
  \textbf{79}(5), 729 (1999)

\bibitem{Lundberg2007}
M.~Lundberg, K.~Krishan, N.~Xu, C.S. O'Hern, M.~Dennin, arXiv:0707.4014v2

\bibitem{was75}
K.~Washizu, \emph{Variational methods in elasticity and plasticity} (Pergamon
  Press, 1975)

\bibitem{vin06}
S.~Vincent-Bonnieu, R.~H\"ohler, S.~Cohen-Addad, preprint

\bibitem{Mason1996}
T.G. Mason, J.~Bibette, D.A. Weitz, J. Colloid Interface Sci. \textbf{179}, 439
  (1996)

\bibitem{Mason1995}
T.G. Mason, J.~Bibette, D.A. Weitz, Phys. Rev. Lett. \textbf{75}, 10 (1995)

\end{thebibliography}

\end{document}